\documentclass[11pt]{article}
\usepackage{amssymb}
\usepackage{amsmath,array,slashed}
\usepackage{amsthm}
\usepackage{epsfig,graphics}
\usepackage[pdftex]{color}
\usepackage{simplemargins}
\usepackage{graphicx}

\setleftmargin{1in}
\setrightmargin{1in}
\settopmargin{1in}
\setbottommargin{1in}

\newcommand{\inp}[1]{\langle #1 \rangle}
\newcommand{\Dl}{\Delta}

\begin{document}

\begin{titlepage}
\begin{center}

\hfill  DAMTP-2008-106

\vskip 2 cm
{\Large \bf Four-point correlators with higher weight superconformal primaries in the AdS/CFT Correspondence}\\
{\vskip 1cm  
Linda I. Uruchurtu\\}
{\vskip 0.75cm
\small Department of Applied Mathematics and Theoretical Physics\\
\small University of Cambridge\\
\small Wilberforce Road, Cambridge CB3 0WA\\
\small \texttt{liu20@damtp.cam.ac.uk}}\\

\end{center}

\vskip 2 cm
\begin{abstract}
\baselineskip=18pt
The four-point correlation function of two 1/2 BPS primaries of conformal weight $\Delta=2$ and two 1/2-BPS
primaries of conformal weight $\Delta=n$ is calculated in the large $\lambda$, large $N$ limit.  These operators are dual
to Kaluza--Klein supergravity fields $s_k$ with masses $m^2=-4$ and $m^2=n(n-4)$. Given that the
existing  formalism for evaluating sums of products of $SO(6)$ tensors that determine the effective couplings is only suitable for primaries with small conformal dimensions, we make us of an alternative formalism based on harmonic polynomials
introduced by Dolan and Osborn.

We then show that the supergravity lagrangian relevant to the computation is of $\sigma$-model
type (\emph{i.e.}, the four-derivative couplings vanish) and that the final result for the connected
amplitude splits into a free and an interacting part, as expected on general grounds.
\end{abstract}
\end{titlepage}

\section{Introduction}
The AdS/CFT correspondence \cite{Maldacena:1997re, Witten:1998qj, Gubser:1998bc}, the celebrated conjecture relating type IIB strings on $AdS_5\times S^5$ to $\mathcal{N}=4$ super Yang-Mills theory, has received a lot of attention given the possibilities of extracting information from the strongly coupled gauge theory, by means of performing perturbative computations in the gravitational dual. However, this same property has made it difficult to find a way to prove the conjecture in all generality, and one needs to rely in tests restricted to the BPS sector.

In particular, evaluation of four-point correlation functions of BPS operators in tree level supergravity has allowed to check  the correspondence in the limit $N\rightarrow \infty$, large $\lambda$.  Four-point functions are very interesting objects as they are not completely fixed by conformal symmetry, and they can be given an Operator Product Expansion (OPE) interpretation, which is known to encode all the dynamical information of the theory. Moreover, their quantum behaviour is severely restricted due to the existence of a lagragian formulation of $d=4$ SYM, so the predictions on the dynamical piece can be verified by direct computation.

The present availability of the spectrum has limited the calculations to fields arising in the compactification of IIB supergravity on $AdS_5\times S^5$. The standard AdS/CFT dictionary relates the infinite tower of KK scalar excitations originating from the trace of the graviton and the five-form on $S^5$ to $1/2$-BPS operators of $\mathcal{N}=4$ SYM theory. These operators are known to have protected conformal dimensions, two- and three-point functions  \cite{Lee:1998bxa,D'Hoker:1998tz}.  Four-point functions are then the simplest objects which exhibit non-trivial dynamics when going to the strongly coupled regime. Therefore, comparison of results obtained from supergravity with those obtained either from free or perturbative YM often reveal new insights into the behaviour of the theory, while also constituting a probing test for the duality.

Given the technical difficulty associated with evaluating diagrams for generic operators, supergravity induced four-point functions have been studied only for specific examples\footnote{Other known examples involving superconformal descendents can be found in \cite{Liu:1998ty,Uruchurtu:2007kq}.}. The first example in the literature, in which the basic techniques for evaluating amplitudes were developed, was the four-point function of dilaton-axion fields \cite{D'Hoker:1999pj}, whose dual operators belong to the (ultrashort) current multiplet of $\mathcal{N}=4$ SYM. Four-point functions of superconformal primaries followed later since the cubic and quartic couplings are difficult to evaluate \cite{Arutyunov:1999en,Arutyunov:1999fb}. The examples have been restricted to those involving four identical operators with weight $\Delta=2,3,4$ \cite{Arutyunov:2000py,Arutyunov:2002fh,Arutyunov:2003ae}, and the results have shown to have the dynamical structure predicted by the gauge theory and superconformal symmetry. The first example that explored the dynamics in the $t$-channel between massless fields and Kaluza-Klein (KK) excitations was presented in \cite{Berdichevsky:2007xd}, and so far, there are not known computations from supergravity that address fields transforming in generic representations, this is, of the form $[0,n,0]$. 

In this paper we then continue the programme of evaluating new examples of four-point functions involving BPS operators. In this case we will consider two operators of lowest conformal dimension $\Delta=2$, and two operators of generic conformal dimension $\Delta=n$. This example generalises the result in \cite{Berdichevsky:2007xd} and is the first one involving operators transforming in generic representations of the $R$-symmetry group. This constitutes a first step towards computing the four-point function of 1/2-BPS primaries of arbitrary weight, while also allowing the emergence of interactions between the massless graviton multiplet and the infinite tower of KK excitations. We will start by establishing the general structure of the amplitude by restricting the functional dependence using superconformal symmetry and the dynamical procedure known as the insertion procedure. We then evaluate the amplitude in AdS supergravity and compare this result against the predictions made in the gauge theory side. 

To this end, one needs to obtain the on-shell value of the five-dimensional effective action for type IIB supergravity on $AdS_5\times S^5$ relevant for the calculation. These terms can be found in \cite{Lee:1998bxa,Arutyunov:1999en,Arutyunov:1999fb, Arutyunov:1998hf}. To calculate the on-shell action, we use the techniques in \cite{Arutyunov:2002fh, Berdichevsky:2007xd, D'Hoker:1999ni} for evaluating the AdS $z$-integrals. However, for the evaluation of the effective vertices coming from the integrals over the $S^5$, we introduce a new method, as the direct evaluation of sums of products of $SO(6)$ $C$-tensors cannot be evaluated in a closed form when including representations depending on generic values\footnote{And even in cases in which $n>4$ it becomes very involved and one requires the use of a computer algebra program.}\cite{Arutyunov:2002fh}. We then show that as in the previous cases in the literature, the four derivative terms in the effective lagrangian can be re-expressed in terms of two and zero derivative terms, so the lagrangian is of $\sigma$-model type. We also show how the resulting quartic lagrangian has a rather simple form, after the dramatic simplification coming from adding the different contributions. Finally, we will verify that the result for the strongly coupled four-point amplitude splits into a free and an interacting piece, which has the structure predicted by the insertion procedure \cite{Arutyunov:2002fh,Intriligator:1998ig}. This phenomena has also been observed in all other four-point functions involving superconformal primary operators, and is a highly non--trivial result as there is no argument supporting this splitting in the gravitational theory. This result serves then as further evidence for the AdS/CFT correspondence

The plan of this paper is as follows. In section \ref{sec:structure} we consider the general structure of the four-point amplitude of $1/2$-BPS operators using the different symmetries (i.e. conformal, crossing and $R$-symmetry) and we see that the dependence is contained in four functions of conformal ratios. In section \ref{sec:insertion}, we introduce further constraints on the interacting piece from the insertion procedure, that reduces the number of independent functions from four to one. Section \ref{sec:chiralpsdiffweight} is devoted to the evaluation of the four-point function of interest in the supergravity approximation. Some technical details are postponed to the appendices, including the derivation of the quartic lagrangian and the novel method for computing the effective interaction vertices coming from integrals on $S^5$. In section \ref{sec:verifying} we analyse the supergravity result in the light of the predictions obtained from the CFT side, and verify that indeed, the supergravity-induced amplitude splits into a free and an interacting piece. We also reveal a puzzling result pertaining to one of the coefficient functions entering the amplitude. Finally, section \ref{sec:conclusions} summarises our results and presents some interesting problems that could be addressed in the future.
%
\section{General Structure of the Four-Point Function}
\label{sec:structure}
\begin{figure}[ht]
\begin{center}
\resizebox{60mm}{100mm}{\input{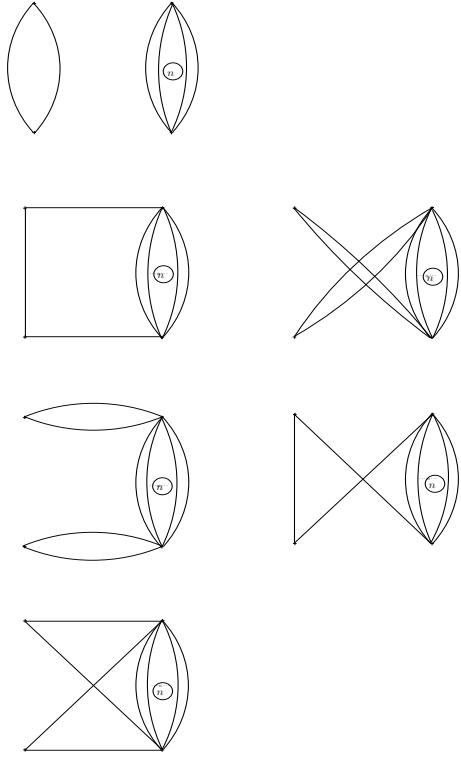}} 
\end{center}
\caption{Propagator basis for the process $\langle \mathcal{O}_{2}(\vec{x}_1) \mathcal{O}_{2}(\vec{x}_2) \mathcal{O}_{n}(\vec{x}_3)\mathcal{O}_{n}(\vec{x}_4)\rangle$. The graphs are arranged in four equivalence classes. The symbol $n$ stands
for the $n$ propagators coming out from the corresponding vertices.}
\label{colourbasis}
\end{figure}

The general structure of the process we are considering is constrained by $R$ and crossing summetry. In this paper we are concerned with four-point functions of $1/2$-BPS superconformal primaries of $\mathcal{N}=4$ supersymmetric Yang-Mills theory. The canonically normalised operators \cite{Lee:1998bxa}  with conformal dimension $\Delta=k$ are given by
\begin{equation}
\mathcal{O}_k^{I}(\vec{x})=\frac{(2\pi)^k}{\sqrt{k\lambda^k}}C_{i_1\cdots i_k}^{I}\mathrm{tr}(\varphi^{i_1}(\vec{x})\cdots \varphi^{i_k}(\vec{x}))
\label{normCPOk}
\end{equation}
where $C_{i_1\cdots i_k}^I$ are totally symmetric traceless $SO(6)$ tensors of rank $k$ and the index $I$ runs over a basis of a representation of $SO(6)$ specified by $k$. The four-point function we wish to study has the form
\begin{equation}
\langle \mathcal{O}^{I_1}_{2}(\vec{x}_1) \mathcal{O}^{I_2}_{2}(\vec{x}_2) \mathcal{O}^{I_3}_{n}(\vec{x}_3) \mathcal{O}^{I_4}_{n}(\vec{x}_4)\rangle
\label{diffweightprocess}
\end{equation}
The content of the OPE's is given by operators in the representations arising in the tensor of the $SU(4)$ representations $[0,2,0]$ and $[0,n,0]$.  This is
\begin{equation}
\label{4pdiffweightreps}
\langle \mathcal{O}_{2}(\vec{x}_1)\mathcal{O}_{2}(\vec{x}_2)\mathcal{O}_{n}(\vec{x}_3)\mathcal{O}_{n}(\vec{x}_4)\rangle
\in [0,2,0] \otimes [0,2,0] \otimes [0,n,0] \otimes [0,n,0]
\end{equation}
where
\begin{equation}
[0,n,0]\otimes[0,n,0]=\sum_{k=0}^{n}\sum_{l=0}^{n-k}[l,2n-2l-2k,l] 
\end{equation}
All the OPE channels with $l=0,1$ contain only short and semishort operators. We now follow the ideas and methods in \cite{Arutyunov:2002fh}. An appropriate basis to study the content of a four-point function is given by the \emph{propagator basis} arising in free field theory. Recall that the propagator for scalar fields is given by
\begin{equation}
\label{scalarpropYM}
\langle \varphi^{i}(\vec{x}_1)\varphi^j(\vec{x}_2)\rangle=\frac{\delta^{ij}}{|\vec{x}_{12}|^2}
\end{equation}
Let us introduce the harmonic (complex) variables $u^i$ satisfying the following constraints
\begin{equation}
u_iu_i=0 \qquad \qquad u_i\bar{u}_i=1
\end{equation}
These variables parametrise the coset $SO(6)/SO(2)\times SO(4)$ so that under an $SO(6)$ transformation, the highest weight vector representation transforms as $u^{i_1}\cdots u^{i_n}$, so projections onto representations $[0,n,0]$ can be achieved by writing
\begin{equation}
\mathcal{O}^{(n)}=u_{i_1}\cdots u_{i_n}\mathrm{tr}(\varphi^{i_1}\cdots \varphi^{i_n})
\end{equation}
with $(n)$ denoting the highest weight of the representation $[0,n,0]$. Scalar fields can also be projected
\begin{equation}
\varphi^{i_1}(\vec{x}_1)=\varphi(1)\bar{u}_1^{i_1}
\end{equation} 
so (\ref{scalarpropYM}) can be rewritten as
\begin{equation}
\langle \varphi(1)\varphi(2) \rangle =\frac{ {u_1}^{i_1} {u_2}^{i_2}\delta^{i_1 i_2}}{|\vec{x}_{12}|^2}
=\frac{ ({u_1}^{i_1} {u_2}^{i_2})}{|\vec{x}_{12}|^2}
\end{equation}
We can now construct four-point functions by connecting pairs of points by propagators. For the case in hand, the amplitude will have $n+2$ contractions, so the propagator basis for (\ref{4pdiffweightreps}) is determined from six graphs belonging to four equivalence classes, as depicted in figure \ref{colourbasis}. 
Each of the propagator structures can be multiplied by an arbitrary function of the conformally invariant ratios $u$ and $v$
\begin{equation}
u=\frac{|\vec{x}_{12}|^2|\vec{x}_{34}|^2}{|\vec{x}_{13}|^2|\vec{x}_{24}|^2} \qquad \qquad
v=\frac{|\vec{x}_{14}|^2|\vec{x}_{23}|^2}{|\vec{x}_{13}|^2|\vec{x}_{24}|^2} 
\label{crossradii}
\end{equation}
Hence, the most general four-point amplitude with the required transformation properties is given by
\begin{eqnarray}
&&\langle \mathcal{O}_{2}(\vec{x}_1) \mathcal{O}_{2}(\vec{x}_2) \mathcal{O}_{n}(\vec{x}_3)\mathcal{O}_{n}(\vec{x}_4)\rangle=
a(u,v)\frac{({u_1}^{i_1}{u_2}^{i_2})^2({u_3}^{i_3}{u_4}^{i_4})^n}{|\vec{x}_{12}|^4|\vec{x}_{34}|^{2n}}
\nonumber \\
&+&b_1(u,v)\frac{({u_1}^{i_1}{u_2}^{i_2})({u_3}^{i_3}{u_4}^{i_4})^{n-1}({u_1}^{i_1}{u_3}^{i_3})({u_2}^{i_2}{u_4}^{i_4})}{|\vec{x}_{12}|^2|\vec{x}_{34}|^{2(n-1)}|\vec{x}_{13}|^2|\vec{x}_{24}|^2}
\nonumber \\
&+&b_2(u,v)\frac{({u_1}^{i_1}{u_2}^{i_2})({u_3}^{i_3}{u_4}^{i_4})^{n-1}({u_1}^{i_1}{u_4}^{i_4})({u_2}^{i_2}{u_3}^{i_3})}{|\vec{x}_{12}|^2|\vec{x}_{34}|^{2(n-1)}|\vec{x}_{14}|^2|\vec{x}_{23}|^2}
\nonumber \\
&+&c_1(u,v)\frac{({u_1}^{i_1}{u_3}^{i_3})^2({u_2}^{i_2}{u_4}^{i_4})^2({u_3}^{i_3}{u_4}^{i_4})^{n-2}}{|\vec{x}_{34}|^{2(n-2)}|\vec{x}_{13}|^{4}|\vec{x}_{24}|^{4}}
+c_2(u,v)\frac{({u_1}^{i_1}{u_4}^{i_4})^2({u_2}^{i_2}{u_3}^{i_3})^2({u_3}^{i_3}{u_4}^{i_4})^{n-2}}{|\vec{x}_{34}|^{2(n-2)}|\vec{x}_{14}|^{4}|\vec{x}_{23}|^{4}}
\nonumber \\
&+&d(u,v)\frac{({u_1}^{i_1}{u_3}^{i_3})({u_2}^{i_2}{u_4}^{i_4})({u_2}^{i_2}{u_3}^{i_3})({u_1}^{i_1}{u_4}^{i_4})({u_3}^{i_3}{u_4}^{i_4})^{n-2}}{|\vec{x}_{13}|^2|\vec{x}_{24}|^{2}|\vec{x}_{23}|^{2}|\vec{x}_{14}|^2|\vec{x}_{34}|^{2(n-2)}}
\label{structure4p}
\end{eqnarray}
Permutation symmetries under exchange of $1\leftrightarrow 2$ and $3 \leftrightarrow 4$ reduce the number of coefficient functions to four since
\begin{eqnarray}
a(u,v)&=&a(u/v,1/v) \nonumber \\
b_2(u,v)&=&b_1(u/v,1/v) \nonumber \\
c_2(u,v)&=&c_1(u/v,1/v) \nonumber \\
d(u,v)&=&d(u/v,1/v)
\end{eqnarray}
The harmonic variables in  (\ref{structure4p}) can be re-expressed in terms of $SO(6)$ $C$-tensors (Appendix \ref{sec:SphereInts}) as
\begin{eqnarray}
&&\langle \mathcal{O}_{2}^{I_1}(\vec{x}_1) \mathcal{O}_{2}^{I_2}(\vec{x}_2) \mathcal{O}_{n}^{I_3}(\vec{x}_3)\mathcal{O}_{n}^{I_4}(\vec{x}_4)\rangle=
a(u,v)\frac{\delta^{I_1 I_2}_2\delta^{I_3 I_4}_n}{|\vec{x}_{12}|^4|\vec{x}_{34}|^{2n}}
+b_1(u,v)\frac{C^{I_1I_2I_3I_4}}{|\vec{x}_{12}|^2|\vec{x}_{34}|^{2(n-1)}|\vec{x}_{13}|^2|\vec{x}_{24}|^2}
\nonumber \\
&+&b_2(u,v)\frac{C^{I_1I_2I_4I_3}}{|\vec{x}_{12}|^2|\vec{x}_{34}|^{2(n-1)}|\vec{x}_{14}|^2|\vec{x}_{23}|^2}
+c_1(u,v)\frac{\Upsilon^{I_1I_2I_3I_4}}{|\vec{x}_{34}|^{2(n-2)}|\vec{x}_{13}|^{4}|\vec{x}_{24}|^{4}}
\nonumber \\
&+&c_2(u,v)\frac{\Upsilon^{I_1I_2I_4I_3}}{|\vec{x}_{34}|^{2(n-2)}|\vec{x}_{14}|^{4}|\vec{x}_{23}|^{4}}
+d(u,v)\frac{S^{I_1I_2I_3I_4}}{|\vec{x}_{13}|^2|\vec{x}_{24}|^{2}|\vec{x}_{23}|^{2}|\vec{x}_{14}|^2|\vec{x}_{34}|^{2(n-2)}}
\label{structure4pCten}
\end{eqnarray}
It is possible to compute the value of the coefficient functions using free field theory in the large $N$ limit (\emph{e.g.} contribution form planar diagrams only). This was done in \cite{Rayson:2007th} and the results are reproduced here\footnote{The coefficient of the disconnected piece is set to be one as a consequence of the normalisation choice for the two-point functions.}
\begin{equation}
a=1 \qquad b_i=\frac{2n}{N^2} \qquad c_i=\frac{n(n-1)}{2N^2}\left(\frac{X_{i_1\cdots i_{n-2}kk}X_{j_1\cdots j_{n-2}ll}}{X_{m_1\cdots m_n}X_{m_1\cdots m_n}}\right) \qquad d=\frac{2n(n-1)}{N^2}
\label{largeNfreecoeff}
\end{equation}
where $X_{i_1\cdots i_n}$ is a totally symmetric rank $n$ colour tensor, so that the value of $c_i$ is dependent on a non-trivial tensor calculation\footnote{For $n=2$, $c_i=1$ and for $n=3$, $c_i=0$. For $n\geq 4$ it was shown in \cite{Rayson:2007th} that it the value of $c_i$ can be approximated as
\begin{equation}
c_i\simeq \frac{2n(n-2)}{N^2}\simeq (n-2)b_i 
\end{equation}
}. Notice also that $d=(n-1)b_i$ for any value of $n$ and $N$.  
\section{The Insertion Formula}
\label{sec:insertion}
We now follow the ideas developed in \cite{Arutyunov:2002fh} to restrict the dynamical piece of the four-point function. The derivative with respect to the coupling $g_{YM}^2$ of the amplitude (\ref{diffweightprocess}) can be expressed as (see also \cite{Intriligator:1998ig})
\begin{equation}
\frac{\partial}{\partial g_{YM}^2}\langle \mathcal{O}_{2}\mathcal{O}_{2}\mathcal{O}_{n}\mathcal{O}_{n}\rangle 
\propto \int d^{4}\vec{x_0} d^{4}\theta_0
\langle \mathcal{O}_\tau(\vec{x_0})\mathcal{O}_{2}\mathcal{O}_{2}\mathcal{O}_{n}\mathcal{O}_{n}\rangle 
\end{equation}
The integration above is consistent with supersymmetry as the $\theta$-expansion for the case $\mathcal{O}_2$ terminates at four $\theta$'s, and one can show that the five-point function in the right side of the previous expression, gives rise to a nilpotent superconformal covariant. By following this procedure in which we insert and additional \emph{ultrashort} operator, it is possible to extract more information about the four-point function we have been studying. As the construction of nilpotents covariants if of technical nature, we refer to \cite{Arutyunov:2002fh} for references and the derivation of the results reproduced below.

The key idea is to assume that the nilpotent covariant must have the following form
\begin{equation}
\label{inserting}
\langle \mathcal{O}_\tau(\vec{x_0})\mathcal{O}_{2}\mathcal{O}_{2}\mathcal{O}_{n}\mathcal{O}_{n}\rangle=R^{2222}(\theta_0)^4F^{00n-2n-2}(\vec{x_0},\cdots,\vec{x}_4,u_1,\cdots,u_4)
\end{equation}
so the five-point function is factorised into a kernel with weight $2$ and an additional factor carrying the remaining $SO(6)$ quantum numbers, so at each point the weight is $k_i'=k_i-2$.  Note here that the Grassmann factor $(\theta_0)^4$ carries the full harmonic dependence at the insertion point. The relevant expressions are given by
\begin{eqnarray}
\label{Rkernel}
R^{2222}&=&u \frac{({u_1}^{i_1}{u_2}^{i_2})^2({u_3}^{i_3}{u_4}^{i_4})^2}{|\vec{x}_{12}|^2|\vec{x}_{34}|^2}+(v-u-1)\frac{({u_1}^{i_1}{u_2}^{i_2})({u_3}^{i_3}{u_4}^{i_4})({u_1}^{i_1}{u_3}^{i_3})({u_2}^{i_2}{u_4}^{i_4})}{|\vec{x}_{12}|^2|\vec{x}_{34}|^2|\vec{x}_{13}|^2|\vec{x}_{24}|^2}
\nonumber \\
&+&(1-u-v)
\frac{({u_1}^{i_1}{u_2}^{i_2})({u_3}^{i_3}{u_4}^{i_4})({u_1}^{i_1}{u_4}^{i_4})({u_2}^{i_2}{u_3}^{i_3})}{|\vec{x}_{12}|^2|\vec{x}_{34}|^2|\vec{x}_{14}|^2|\vec{x}_{23}|^2}+
 \frac{({u_1}^{i_1}{u_3}^{i_3})^2({u_2}^{i_2}{u_4}^{i_4})^2}{|\vec{x}_{13}|^4|\vec{x}_{24}|^4}
 \nonumber \\
&+& \frac{({u_1}^{i_1}{u_4}^{i_4})^2({u_2}^{i_2}{u_3}^{i_3})^2}{|\vec{x}_{14}|^4|\vec{x}_{23}|^4}+(u-v-1)\frac{({u_1}^{i_1}{u_3}^{i_3})({u_1}^{i_1}{u_4}^{i_4})({u_2}^{i_2}{u_4}^{i_4})({u_2}^{i_2}{u_3}^{i_3})}{|\vec{x}_{13}|^2|\vec{x}_{14}|^2|\vec{x}_{23}|^2|\vec{x}_{24}|^2}
\end{eqnarray}
and
\begin{equation}
F^{0 k_1'k_2'k_3'}=
\left(\frac{{u_2}^{i_2}{u_3}^{i_3}}{|\vec{x}_{23}|^2}\right)^{\frac{1}{2}(k_1+k_2-k_3-2)}
\left(\frac{{u_2}^{i_2}{u_4}^{i_4}}{|\vec{x}_{24}|^2}\right)^{\frac{1}{2}(k_1+k_3-k_2-2)}
\left(\frac{{u_3}^{i_3}{u_4}^{i_4}}{|\vec{x}_{34}|^2}\right)^{\frac{1}{2}(k_2+k_3-k_1-2)}f(\vec{x}_0,\cdots, \vec{x}_4)
\end{equation}
Substitution of these expressions into (\ref{inserting}) and integration over the Grassman variable $\theta_0$  lead to the following dependence on the coupling of the four-point function (\ref{diffweightprocess})
\begin{eqnarray}
\label{insertionres}
&&\frac{\partial}{\partial g_{YM}^2}\langle \mathcal{O}_{2}^{I_1}\mathcal{O}_{2}^{I_2}\mathcal{O}_{n}^{I_3}\mathcal{O}_{n}^{I_4}\rangle=u G(u,v)\frac{\delta_{2}^{I_1I_2}\delta_{n}^{I_3I_4}}{|\vec{x}_{12}|^{4}|\vec{x}_{34}|^{2n}}+
(v-u-1)G(u,v)\frac{C^{I_1I_2I_3I_4}}{|\vec{x}_{12}|^2|\vec{x}_{34}|^{2(n-1)}|\vec{x}_{13}|^2|\vec{x}_{24}|^2}
\nonumber \\
&&+(1-u-v)G(u,v)\frac{C^{I_1I_2I_4I_3}}{|\vec{x}_{12}|^2|\vec{x}_{34}|^{2(n-1)}|\vec{x}_{14}|^2|\vec{x}_{23}|^2}+G(u,v)\frac{\Upsilon^{I_1I_2I_3I_4}}{|\vec{x}_{34}|^{2(n-2)}|\vec{x}_{13}|^{4}|\vec{x}_{24}|^{4}}
\nonumber \\
&&+vG(u,v)\frac{\Upsilon^{I_1I_2I_4I_3}}{|\vec{x}_{34}|^{2(n-2)}|\vec{x}_{14}|^{4}|\vec{x}_{23}|^{4}}+(u-v-1)G(u,v)\frac{S^{I_1I_2I_3I_4}}{|\vec{x}_{13}|^2|\vec{x}_{24}|^{2}|\vec{x}_{23}|^{2}|\vec{x}_{14}|^2|\vec{x}_{34}|^{2(n-2)}}\nonumber \\
\end{eqnarray}
with
\begin{equation}
G(u,v)=\int d^4 \vec{x}_0 f(\vec{x}_0,\cdots, \vec{x}_4)
\end{equation}
So comparing (\ref{insertionres}) with (\ref{structure4pCten}) one realises that the amplitude depends on a single function $\mathcal{F}(u,v)$, satisfying
\begin{eqnarray}
a(u,v)&=&u\mathcal{F}(u,v) \nonumber \\
b_1(u,v)&=&(v-u-1)\mathcal{F}(u,v)\nonumber \\
b_2(u,v)&=&(1-u-v)\mathcal{F}(u,v)\nonumber \\
c_1(u,v)&=&\mathcal{F}(u,v)\nonumber \\
c_2(u,v)&=&v\mathcal{F}(u,v)\nonumber \\
d(u,v)&=&(u-v-1)\mathcal{F}(u,v)
\label{insertrels}
\end{eqnarray}
This is a (partial) non-renormalisation theorem for the structure of the amplitude (i.e. a dynamical constraint), so verification of this result from the supergravity calculation constitutes an indirect test for the AdS/CFT correspondence.
\section{Supergravity Calculation}
\label{sec:chiralpsdiffweight}
The precise relation between the operators in the gauge theory and the fields in the bulk was established in \cite{Witten:1998qj,Gubser:1998bc} and refined in \cite{Mueck:1999kk,Mueck:1999gi,Bena:1999jv}. The proposition is 
 \begin{equation}
\langle \exp\{ \int d^{4}x \phi_{0}(\vec{x})\mathcal{O}(\vec{x})\}\rangle_{CFT}=\exp \{ -S_{IIB}[\phi_{0}(\vec{x})] \}
\label{prescriptionadscft}
\end{equation}
On the left hand side of (\ref{prescriptionadscft}) the field $\phi_0(\vec{x})$, which stands for the boundary value of the bulk field $\phi(z_0,\vec{x})$, is a source for the operator $\mathcal{O}(\vec{x})$, and the expectation value is computed by expanding the exponential and evaluating the correlation functions in the field theory. On the right hand side, one has the generating functional encompassing all dynamical processes of IIB strings on $AdS_5\times S^{5}$. In the supergravity approximation, $S_{IIB}$ is just the type IIB supergravity action on $AdS_5\times S^5$, and it is assumed here that all the bulk fields $\phi(z_0,\vec{x})$ have appropriate boundary behaviour so they source the YM operators on the left hand side. Hence in practice, one first finds the boundary data for the corresponding gravitational fields and then computes correlation functions as a function of these values (on-shell), by functional differentiation.

Given that we are interested in computing correlation functions of superconformal primaries, we first need to identify the bulk fields whose value in the boundary serve as sources.  From looking at the representations, we see that the fields dual to superconformal primaries are obtained from mixtures of modes from the graviton and the five form on the $S^5$ \cite{Kim:1985ez} and are denoted as $s_k^I$, with $I$ running over the basis of the corresponding $SO(6)$ irrep. with Dynkin labels $[0,k,0]$. The four-point function can then be determined from the expression
\begin{equation}
\langle \mathcal{O}^{I_1}_{k_1}(\vec{x}_{1})\mathcal{O}^{I_2}_{k_2}(\vec{x}_{2})\mathcal{O}^{I_3}_{k_3}(\vec{x}_{3})\mathcal{O}^{I_4}_{k_4}(\vec{x}_{4})\rangle=
\frac{\delta}{\delta s_{k_1}^{I_{1}}(\vec{x}_{1})}\frac{\delta}{\delta s_{k_2}^{I_{2}}(\vec{x}_{2})}
\frac{\delta}{\delta s_{k_3}^{I_{3}}(\vec{x}_{3})}\frac{\delta}{\delta s_{k_4}^{I_{4}}(\vec{x}_{4})}(-S_{IIB})
\end{equation}
\subsection{On-Shell Lagrangian}
\label{subsec:onshelldiffCPO}
We are interested in computing (\ref{diffweightprocess}) in strongly coupled $\mathcal{N}=4$ SYM theory, using the supergravity approximation. The prescription (\ref{prescriptionadscft}) indicates that we need to evaluate the on-shell value of the five-dimensional effective action of compactified type IIB supergravity on $AdS_5\times S^5$. We write this action as
\begin{equation}
S=\frac{N}{8\pi^2}\int [dz] \left(\mathcal{L}_2+\mathcal{L}_3+\mathcal{L}_4\right)
\end{equation}
which involves the sum of quadratic, cubic and quartic terms. The normalisation of the action can be derived from expressing the ten dimensional gravitational coupling as $2\kappa_{10}^2=(2\pi)^7g_s^2\alpha'^4$ and using the volume of $S^5$ to get the five dimensional gravitational coupling
\begin{equation}
\label{normaction5d}
\frac{1}{2\kappa_{5}^2}=\frac{\mathrm{Vol}(S^5)}{2\kappa_{10}^2}=\frac{N^2}{8\pi^2l^3}
\end{equation}
with $l$ being the $AdS_5$ radius, which will be set to one. The quadratic terms \cite{Kim:1985ez,Arutyunov:1998hf} read
\begin{eqnarray}
\mathcal{L}_{2}&=&\frac{1}{4}(D_{\mu}{s_{2}}^{1}D^{\mu}{s_{2}}^{1}-4{s_{2}}^{1}{s_{2}}^{1})
               +\frac{1}{4}(D_{\mu}{s_{n}}^{1}D^{\mu}{s_{n}}^{1}+n(n-4){s_{n}}^{1}{s_{n}}^{1})
               \nonumber \\
               &+&\frac{1}{2}({F_{\mu \nu,1}}^{1})^{2} +\frac{1}{2}(({F_{\mu \nu,n-1}}^{1})^{2}+2n(n-2)(A^{1}_{\mu,n-1})^{2})
               \nonumber \\
               &+&\frac{1}{4}D_{\mu}\phi_{\nu \rho,0}D^{\mu}\phi_{0}^{\nu \rho}-\frac{1}{2}D_{\mu}\phi^{\mu \nu,0}D^{\rho}\phi_{\rho \nu,0}
               +\frac{1}{2}D_{\mu}\phi^{\nu}_{\nu,0}D_{\rho}\phi^{\mu \rho}_{0}-\frac{1}{4}D_{\mu}\phi^{\nu}_{\nu,0}D^{\mu}\phi^{\rho}_{\rho,0}
               \nonumber \\
               &-&\frac{1}{2}\phi_{\mu \nu,0}\phi^{\mu \nu}_{0}+\frac{1}{2}(\phi^{\mu}_{\mu,0})^{2}
               \nonumber \\
               &+&\frac{1}{4}D_{\mu}\phi_{\nu \rho,n-2}D^{\mu}\phi_{n-2}^{\nu \rho}-\frac{1}{2}D_{\mu}\phi^{\mu \nu,n-2}D^{\rho}\phi_{\rho \nu,n-2}
               +\frac{1}{2}D_{\mu}\phi^{\nu}_{\nu,n-2}D_{\rho}\phi^{\mu \rho}_{n-2}
               \nonumber \\
               &-&\frac{1}{4}D_{\mu}\phi^{\nu}_{\nu,n-2}D^{\mu}\phi^{\rho}_{\rho,n-2}
               +\frac{(n^2-6)}{4}\phi_{\mu \nu,n-2}\phi^{\mu \nu}_{n-2}-\frac{(n^2-2)}{4}(\phi^{\mu}_{\mu,n-2})^{2}
\end{eqnarray}
where $F_{\mu \nu,k}=\partial_{\mu}A_{\nu,k}-\partial_{\nu}A_{\mu,k}$, and summation over upper indices is assumed, running over the basis of the irreducible representation corresponding to the field\footnote{We often use the notation $s_k^{I_m}\equiv s_{k}^{m}$.}. We should point out that the fields have been rescaled in order to simplify the action. In this case, the corresponding rescaling factors are given by
\begin{equation}
s_{n} \rightarrow \sqrt{ \frac{(n+1)}{2^{6}n(n-1)(n+2)}} s_{n} \qquad A_{\mu,n-1} \rightarrow
2\sqrt{\frac{n+1}{n}} A_{\mu,n-1}
\end{equation}
and all symmetric tensors are left unscaled. The cubic couplings \cite{Arutyunov:1999en, Lee:1998bxa, Lee:1999pj} are given by
\begin{eqnarray}
\mathcal{L}_{3}&=&-\frac{1}{3}\langle C^{1}_{2}C^{2}_{2}C^{3}_{[0,2,0]}\rangle s_{2}^{1}s_{2}^{2}s_{2}^{3}
                -\frac{n(n-1)}{2}\langle C^{1}_{n}C^{2}_{n}C^{3}_{[0,2,0]} \rangle s_{n}^{1}s_{n}^{2}s_{2}^{3}
                \nonumber \\
               &-&\frac{1}{4}\left(D^{\mu}s_{2}^{1}D^{\nu}s_{2}^{1}\phi_{\mu \nu,0}-\frac{1}{2}(D^{\mu}s_{2}^{1}D_{\mu}s_{2}^{1}
               -4s_{2}^{1}s_{2}^{1})\phi^{\nu}_{\nu,0} \right)
               \nonumber \\
               &-&\frac{1}{4}\left(D^{\mu}s_{n}^{1}D^{\nu}s_{n}^{1}\phi_{\mu \nu,0}-\frac{1}{2}(D^{\mu}s_{n}^{1}D_{\mu}s_{n}^{1}
               +n(n-4)s_{n}^{1}s_{n}^{1})\phi^{\nu}_{\nu,0} \right)
               \nonumber \\
               &-&\frac{1}{2}\langle C_{2}^{1}C_{n}^{1}C^{3}_{[0,n-2,0]}\rangle \left(D^{\mu}s_{2}^{1}D^{\nu}s_{n}^{1}\phi_{\mu \nu,n-2}
               -\frac{1}{2}(D^{\mu}s_{2}^{1}D_{\mu}s_{n}^{1}-2n s_{2}^{1}s_{n}^{1})\phi^{\nu}_{\nu,n-2} \right)
               \nonumber \\
               &-&\langle C^{1}_{2}C^{2}_{2}C^{3}_{[1,0,1]}\rangle s_{2}^{1}D^{\mu}s_{2}^{2}A_{\mu,1}^{3}
               -\frac{n}{2}\langle C^{1}_{n}C^{2}_{n}C^{3}_{[1,0,1]}\rangle s_{n}^{1}D^{\mu}s_{n}^{2}A_{\mu,1}^{3}
               \nonumber \\
               &-&\sqrt{\frac{n(n-1)}{2}}\langle C^{1}_{2}C^{2}_{n}C^{3}_{[1,n-2,1]}\rangle s_{2}^{1}D^{\mu}s_{n}^{2}A_{\mu,n-1}^{3}
               -\sqrt{\frac{n(n-1)}{2}}\langle C^{1}_{n}C^{2}_{2}C^{3}_{[1,n-2,1]}\rangle s_{n}^{1}D^{\mu}s_{2}^{2}A_{\mu,n-1}^{3}
               \nonumber
\end{eqnarray}
As one can see, there are different contributions to the $s$ and $t$-channels. Finally, the quartic couplings are given by
\begin{equation}
\mathcal{L}_{4}=\mathcal{L}_{4}^{(0)}+\mathcal{L}_{4}^{(2)}+\mathcal{L}_{4}^{(4)}
\end{equation}
where the supraindex indicates contributions coming from zero, two and four-derivative terms, which are given by
\begin{eqnarray}
\mathcal{L}_{4}&=&\mathcal{L}_{k_{1}k_{2}k_{3}k_{4}}^{(0)I_{1}I_{2}I_{3}I_{4}}s_{k_{1}}^{I_{1}}s_{k_{2}}^{I_{2}}s_{k_{3}}^{I_{3}}s_{k_{4}}^{I_{4}}+
\mathcal{L}_{k_{1}k_{2}k_{3}k_{4}}^{(2)I_{1}I_{2}I_{3}I_{4}}s_{k_{1}}^{I_{1}}D_{\mu}s_{k_{2}}^{I_{2}}s_{k_{3}}^{I_{3}}D^{\mu}s_{k_{4}}^{I_{4}}
\nonumber \\
&+&\mathcal{L}_{k_{1}k_{2}k_{3}k_{4}}^{(4)I_{1}I_{2}I_{3}I_{4}}s_{k_{1}}^{I_{1}}D_{\mu}s_{k_{2}}^{I_{2}}D^{\nu}D_{\nu}(s_{k_{3}}^{I_{3}}D^{\mu}s_{k_{4}}^{I_{4}})
\end{eqnarray}
The explicit form of these terms has been computed in \cite{Arutyunov:1999fb}. 
For our case, two of the $k_{i}$'s are equal to 2 and the other two are equal to $n$. This allows for six possible permutations, where the indices $I_{i}$ run over the basis of the representation $[0,k_{i},0]$ which is being summed over. The less trivial part of the calculation is to compute the explicit coefficients of these terms. It can be shown, however, that the relevant interactions can be reduced to a simple expression, as it occurs in all the examples that have been computed previously. We refer to appendix \ref{sec:QuarticSimp} for the details, and reproduce the final expression here
\begin{eqnarray}
\mathcal{L}_{4}&=&-\frac{1}{4}(C^{1234}+S^{1234})s_{2}^{1}D_{\mu}s_{2}^{2}s_{n}^{3}D^{\mu}s_{n}^{4}
\nonumber \\
&+&\frac{1}{8}n(-\delta^{12}_{2}\delta^{34}_{n}+(6+n)C^{1234}+(3n-4)S^{1234}-n\Upsilon^{1234})s_{2}^{1}s_{2}^{1}s_{n}^{3}s_{n}^{4}
\end{eqnarray}
which can be shown to reproduce the $n=3$ case in \cite{Berdichevsky:2007xd}. The quantities in this expression will be defined later. It should be noted that all four derivative terms disappear, which is consistent with the fact that this is a sub-subextremal process, i.e. $k_{1}=k_{2}+k_{3}+k_{4}-4$, as indicated in \cite{Arutyunov:2000ima,D'Hoker:2000dm}.

\begin{figure}[ht]
\begin{center}
\resizebox{120mm}{40mm}{\input{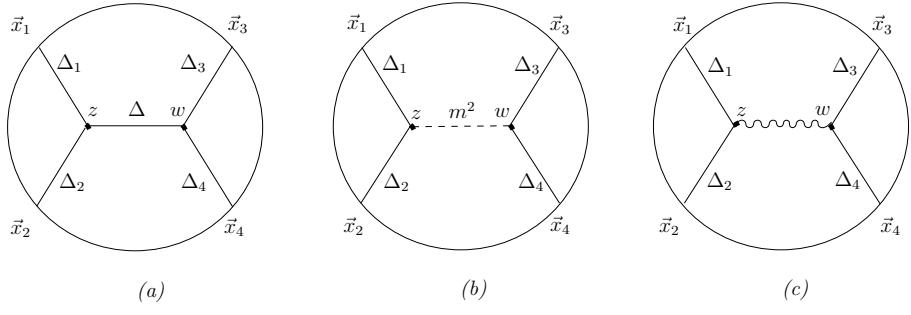}} 
\end{center}
\caption{Witten Diagrams for the $s$-channel process. \emph{(a)} exchange by a scalar with $m^2=-4$ \emph{(b)} exchange by a massless vector \emph{(c)} graviton exchange}
\label{schanneldiffw}
\end{figure}

\begin{figure}[ht]
\begin{center}
\resizebox{160mm}{40mm}{\input{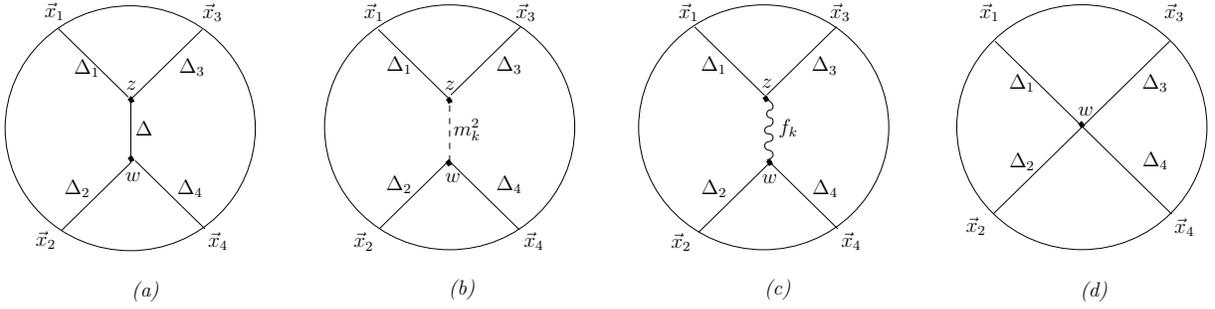}} 
\end{center}
\caption{Witten Diagrams for the $t$-channel process. \emph{(a)} exchange by a scalar of mass $m^2=\Delta(\Delta-4)$ \emph{(b)} exchange by a vector of mass $m_{k}^2=k^2-1$ \emph{(c)} exchange by a tensor field of mass $f_k=k(k+4)$ \emph{(d)} Contact diagram. \ }
\label{tchanneldiffw}
\end{figure}

Now that the relevant terms in the lagrangian have been specified, it remains to compute its on-shell value. From the couplings, one can determine the diagrams that need to be computed. In the $s$-channel, one has a scalar exchange of $s_{2}^{I}$, a vector exchange $A^{I}_{a,[1,0,1]}$ and a graviton exchange, $\phi_{ab,[0,0,0]}$. In the $t$-channel, one has a scalar exchange of $s_{n}^{I}$, a vector exchange $A^{I}_{a,[1,n-1,1]}$ and a massive symmetric tensor $\phi_{ab,[0,n-2,0]}$. Finally one has contact diagrams contributing to the process. The Witten diagrams for the $s$-channel are shown on Fig. \ref{schanneldiffw}. The corresponding diagrams for the $t$-channel and the contact diagram are shown on Fig. \ref{tchanneldiffw}.

It is convenient to introduce the currents
\begin{eqnarray}
T_{\mu\nu}&=&D_{(\mu}s_{k_{1}}D_{\nu)}s_{k_{1}}-\frac{1}{2}g_{\mu\nu}\left(D^{\rho}s_{k_{1}}D_{\rho}s_{k_{2}}+\frac{1}{2}(m^2_{k_{1}}+m^2_{k_{2}}-k_{3}(k_{3}+4))s_{k_{1}}s_{k_{2}}\right) \nonumber \\
J_{\mu}&=&s_{k_{1}}D_{\mu}s_{k_{2}}-s_{k_{2}}D_{\mu}s_{k_{1}}
\end{eqnarray}
where $k_{1},k_{2},k_{3}$ are the conformal weights of the corresponding scalar operators and the primaries here have the appropriate weight depending of the channel one is considering. One then represents the solution to the equations of motion in the form
\begin{equation}
s_{k}=s_{k}^{0}+\tilde{s}_{k} \qquad A_{\mu}=A_{\mu}^{0}+\tilde{A}_{\mu} \qquad \phi_{\mu\nu}=\phi_{\mu\nu}^{0}+\tilde{\phi}_{\mu\nu}
\end{equation}
where $s^{0}_{k}$, $A_{\mu}^{0}$ and $\phi_{\mu\nu}^{0}$ are solutions to the linearised equations with fixed boundary conditions and $\tilde{s}_{k}$, $\tilde{A}_{\mu}$ and $\tilde{\phi}_{\mu\nu}$ represent the fields in the AdS bulk with vanishing boundary conditions. It is then possible to express these fields in terms of an integral on the bulk, involving the corresponding Green function. For the $s$-channel process one needs
\begin{eqnarray}
\tilde{s}_{2}^{5}(w)&=&2\langle C^{1}_{2}C^{2}_{2}C^{5}_{[0,2,0]}\rangle \int [dz] G_{2}(z,w) s^{1}_{2}(z)s^{2}_{2}(z) +n(n-1)\langle C^{1}_{2}C^{2}_{n}C^{5}_{[0,n,0]}\rangle\int [dz] G_{n}(z,w)s^{1}_{2}(z)s^{2}_{n}(z) 
\nonumber \\
\tilde{A}_{\mu,1}^{5}(w)&=&\frac{1}{4}\langle C^{1}_{2}C^{2}_{2}C^{5}_{[1,0,1]} \rangle \int [dz] {G_{\mu}}^{\nu}(z,w)J_{\nu}(z) 
\nonumber \\
\tilde{\phi}^{5}_{\mu\nu,0}(w)&=&
\frac{1}{4}\langle C^{1}_{2}C^{3}_{2}C^{5}_{[0,0,0]} \rangle\int [dz]G_{\mu\nu\mu'\nu'}(z,w)T^{\mu'\nu'}(z)
\end{eqnarray}
where the $z$-integral is being done on the vertex involving the $\mathcal{O}_{2}$'s. For the $t$-channel process, the bulk fields couple to a $\Delta=2$ primary and a $\Delta=n$ primary, so the $z$-integrals read
\begin{eqnarray}
\tilde{s}_{n}^{5}(w)&=&2n(n-1)\langle C^{1}_{2}C^{3}_{n}C^{5}_{[0,n,0]}\rangle \int [dz] G_{n}(z,w) s^{1}_{2}(z)s^{3}_{n}(z) 
\nonumber \\
\tilde{A}_{\mu,n-1}^{5}(w)&=&\frac{1}{2}\sqrt{\frac{n(n-1)}{2}}\langle C^{1}_{2}C^{3}_{n}C^{5}_{[1,n-2,1]} \rangle \int [dz] {G_{\mu}}^{\nu}(z,w)J_{\nu}(z) 
\nonumber \\
\tilde{\phi}^{5}_{\mu\nu,n-2}(w)&=&\frac{1}{2}
\langle C^{1}_{2}C^{3}_{n}C^{5}_{[0,n-2,0]} \rangle \int [dz]G_{\mu\nu\mu'\nu'}(z,w)T^{\mu'\nu'}(z)
\end{eqnarray}
and the currents are defined with the appropriate weights. We will drop the tilde in the following. Using the expressions above, we arrive at the following expression for the on-shell value of the action for each of the channels we are considering. For the $s$-channel, the amplitude is determined by
\begin{eqnarray}
\mathcal{L}_{s-channel}&=&-n(n-1)
\langle C_{2}^{1}C_{2}^{2}C_{2}^{5}\rangle \langle C_{n}^{3}C_{n}^{4}C_{2}^{5}\rangle
\int [dz] s^{1}_{2}(z)s^{2}_{2}(z)G(z,w)s^{3}_{n}(w)s^{4}_{n}(w) \nonumber \\
&-&\frac{n}{2^4}\langle C^{1}_{2}C^{2}_{2}C^{5}_{[1,0,1]} \rangle\langle C^{3}_{2}C^{2}_{4}C^{5}_{[1,0,1]} \rangle  \int [dz]J^{\mu}(z)G_{\mu\nu}(z,w)J^{\nu}(w) \nonumber \\
&-&\frac{1}{2^4}\langle C_{2}^{1}C_{2}^{2}C_{[0,0,0]}^{5}\rangle \langle C_{n}^{3}C_{n}^{4}C_{[0,0,0]}^{5}\rangle
\int [dz] T^{\mu\nu}_{22}(z)G_{\mu\nu\mu'\nu'}(z,w)T^{\mu'\nu'}_{nn}(w) \label{s-channel}
\end{eqnarray}
and for the $t$-channel one has
\begin{eqnarray}
\mathcal{L}_{t-channel}&=&-n^2(n-1)^2
\langle C_{2}^{1}C_{n}^{3}C_{n}^{5}\rangle \langle C_{2}^{2}C_{n}^{4}C_{n}^{5}\rangle
\int [dz] s^{1}_{2}(z)s^{3}_{n}(z)G(z,w)s^{2}_{2}(w)s^{4}_{n}(w) 
\nonumber \\
&-&\frac{n(n-1)}{2^3}\langle C^{1}_{2}C^{3}_{n}C^{5}_{[1,n-2,1]} \rangle\langle C^{2}_{2}C^{4}_{n}C^{5}_{[1,n-2,1]} \rangle \int [dz]J^{\mu}(z)G_{\mu\nu}(z,w)J^{\nu}(w) 
\nonumber \\
&-&\frac{1}{2^3}
\langle C_{2}^{1}C_{n}^{3}C_{[0,n-2,0]}^{5}\rangle \langle C_{2}^{2}C_{n}^{4}C_{[0,n-2,0]}^{5}\rangle
\int [dz] T^{\mu\nu}_{2n}(z)G_{\mu\nu\mu'\nu'}(z,w)T^{\mu'\nu'}_{2n}(w) \label{t-channel}
\end{eqnarray}
The expressions in brackets arise from the integrals over $S^5$ and are defined in appendix \ref{sec:SphereInts}. We will worry about contact interactions later. So far, we see that we need to compute three Witten Diagrams for each channel, involving exchanges of scalars, massless and massive gauge bosons and massless and massive gravitons. In order to do so, we extend the methods developed in \cite{D'Hoker:1999pj,D'Hoker:1999ni,Berdichevsky:2007xd} to perform the computations.
\subsection{Results for Exchange Integrals}
\begin{figure}[t]
\begin{center}
\resizebox{60mm}{38mm}{\input{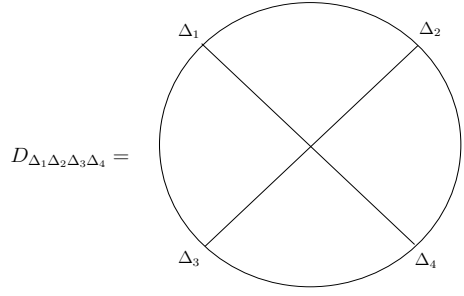}} 
\end{center}
\caption{Graphic representation of a $D$-function.}
\label{wittendfunc}
\end{figure}

We now carry out the integrals and write the results in terms of $\bar{D}$-functions, which are functions of $u$ and $v$ and are related to the more familiar $D$-functions \cite{D'Hoker:1999pj} which are defined as
\begin{equation}
D_{\Delta_1\Delta_2\Delta_3,\Delta_4}(\vec{x}_1,\vec{x}_2,\vec{x}_3,\vec{x}_4)=\int [dw] \tilde{K}_{\Delta_1}(w,\vec{x}_1)\tilde{K}_{\Delta_2}(w,\vec{x}_2)\tilde{K}_{\Delta_3}(w,\vec{x}_3)\tilde{K}_{\Delta_4}(w,\vec{x}_4) 
\end{equation}
where $\tilde{K}_{\Delta}(w,\vec{x})$ is the unit normalised bulk-to-boundary propagator for a scalar of conformal dimension $\Delta$
\begin{equation}
\tilde{K}_{\Delta}(z,\vec{x})=\left(\frac{z_{0}}{z_{0}^2+(\vec{z}-\vec{x})^2}\right)^{\Delta}
\end{equation}
$D_{\Delta_1\Delta_2\Delta_3\Delta_4}$ can be identified as a quartic scalar interactions (see Fig. \ref{wittendfunc}). The relation between the $D$-functions and the $\bar{D}$-functions, and their properties can be found in appendix \ref{sec:Dfunc}. 
Let us first introduce the following notation for the various exchange integrals that contribute to the amplitude. 
\begin{eqnarray}
S_{\Dl_{1}\Dl_{2}\Dl_{3}\Dl_{4}}(\vec{x}_{1},\vec{x}_{2},\vec{x}_{3},\vec{x}_{4})&=&\int [dw] [dz]\tilde{K}_{\Dl_{1}}(z,\vec{x}_{1})\tilde{K}_{\Dl_{2}}(z,\vec{x}_{2})G(z,w)\tilde{K}_{\Dl_{3}}(w,\vec{x}_{3})\tilde{K}_{\Dl_{4}}(w,\vec{x}_{4})
\nonumber \\
V_{\Dl_{1}\Dl_{2}\Dl_{3}\Dl_{4}}(\vec{x}_{1},\vec{x}_{2},\vec{x}_{3},\vec{x}_{4})&=&\int [dw] [dz]\tilde{K}_{\Dl_{1}}(z,\vec{x}_{1})\buildrel\leftrightarrow\over{D^{\mu}}\tilde{K}_{\Dl_{2}}z,\vec{x}_{2})
G_{\mu\nu}(z,w)\tilde{K}_{\Dl_{3}}(w,\vec{x}_{3})\buildrel\leftrightarrow\over{D^{\nu}}\tilde{K}_{\Dl_{4}}(w,\vec{x}_{4}) 
\nonumber \\
T_{\Dl_{1}\Dl_{2}\Dl_{3}\Dl_{4}}(\vec{x}_{1},\vec{x}_{2},\vec{x}_{3},\vec{x}_{4})&=&\int [dz] [dw] T^{\mu\nu}_{\Dl_{1}\Dl_{2}}(z,\vec{x}_{1},\vec{x}_{2})G_{\mu\nu\mu'\nu'}(z,w)T^{\mu'\nu'}_{\Dl_{3},\Dl_{4}}(w,\vec{x}_{3},\vec{x}_{4}) 
\end{eqnarray}
with the bulk-to-bulk propagators appropriately chosen, depending on the particle that is being exchanged.  For our case, the $s$-channel integrals yield
\begin{eqnarray}
S_{22nn}(\vec{x}_{1},\vec{x}_{2},\vec{x}_{3},\vec{x}_{4})&=&\frac{\pi^2}{8}\frac{1}{(n-1)\Gamma(n)} 
\frac{u}{{|\vec{x}_{12}|}^{4}{|\vec{x}_{34}|}^{2n}}\bar{D}_{11nn}
\nonumber \\
V_{22nn}(\vec{x}_{1},\vec{x}_{2},\vec{x}_{3},\vec{x}_{4})&=&-\frac{\pi^2}{4\Gamma(n)}\frac{u}{{|\vec{x}_{12}|}^{4}{|\vec{x}_{34}|}^{2n}}
\left\{-2\bar{D}_{21nn+1}+\bar{D}_{21n+1n}+\bar{D}_{12nn+1}\right\}
\nonumber \\
T_{22nn}(\vec{x}_{1},\vec{x}_{2},\vec{x}_{3},\vec{x}_{4})&=&-\frac{\pi^2}{2\Gamma(n)}\frac{u}{{|\vec{x}_{12}|}^{4}{|\vec{x}_{34}|}^{2n}}
\left\{\frac{1}{3}n\bar{D}_{11nn}-n(n-1)u\bar{D}_{22nn}\right. 
\nonumber \\
&-&\left.n(1+v-u)\bar{D}_{22n+1n+1}\right\}
\label{schannresults}
\end{eqnarray}
and the $t$-channel amplitudes are given by
\begin{eqnarray}
S_{2n2n}(\vec{x}_{1},\vec{x}_{3},\vec{x}_{2},\vec{x}_{4})&=&\frac{\pi^2}{8}\frac{1}{(n-1)\Gamma(n)}\frac{u^2}{{|\vec{x}_{12}|}^{4}{|\vec{x}_{34}|}^{2n}}\bar{D}_{12n-1n}
\nonumber \\
V_{2n2n}(\vec{x}_{1},\vec{x}_{3},\vec{x}_{2},\vec{x}_{4})&=&-\frac{\pi^2}{2n\Gamma(n)}\frac{u^2}{{|\vec{x}_{12}|}^{4}{|\vec{x}_{34}|}^{2n}}
\left\{-\bar{D}_{31nn}+\bar{D}_{12nn+1}-(n-1)\bar{D}_{22n-1n+1}\right.
\nonumber \\
&+&\left.(n-1)u\bar{D}_{23n-1n}\right\}
\nonumber \\
T_{2n2n}(\vec{x}_{1},\vec{x}_{3},\vec{x}_{2},\vec{x}_{4})&=&-\frac{\pi^2}{\Gamma(n)}\left[\frac{n}{(n+1)(n+2)}\right]\frac{u^2}{{|\vec{x}_{12}|}^{4}{|\vec{x}_{34}|}^{2n}}
\left\{2\bar{D}_{31n+1n+1}\right.
\nonumber \\
&+&+n(n-1)u\bar{D}_{33n-1n+1}+\left.2n(1-v-u)\bar{D}_{23nn+1}\right\}
\label{rchannresults}
\end{eqnarray}
where $u$ and $v$ were introduced in (\ref{crossradii}).
These expressions are to be substituted in the action, including an overall factor of $C(n)^2C(2)^2$ where
\begin{equation}
C(n)=\begin{cases}
              \frac{\Gamma(n)}{\pi^{2}\Gamma(n-2)}, \qquad n> 2 \\
              \frac{1}{\pi^2}, \hspace{16mm} n=2
\end{cases}
\end{equation}
\subsection{Contact Diagrams}
One starts from the quartic lagrangian
\begin{eqnarray}
\mathcal{L}_{4}&=&-\frac{1}{4}(C^{1234}-S^{1234})s_{2}^{1}\nabla_{\mu}s_{2}^{2}s_{n}^{3}\nabla^{\mu}s_{n}^{4}
\nonumber \\
&+&\frac{1}{8}n(-\delta_{2}^{12}\delta_{n}^{34}+(6+n)C^{1234}+(3n-4)S^{1234}
-n\Upsilon^{1234})s_{2}^{1}s_{2}^{1}s_{n}^{3}s_{n}^{4}
\label{quartic}
\end{eqnarray}
We record the useful identity
\begin{equation}
D_{\mu}K_{\Delta_{1}}(z,\vec{x}_{1})D^{\mu}K_{\Delta_{2}}(z,\vec{x}_{2}) =\Delta_{1}\Delta_{2}
(K_{\Delta_{1}}(z,\vec{x}_{1})K_{\Delta_{2}}(z,\vec{x}_{2})-2|\vec{x}_{12}|^{2}K_{\Delta_{1}+1}(z,\vec{x}_{1})K_{\Delta_{2}+1}(z,\vec{x}_{2}))
\end{equation}
Using this expression and the definition of the $D$-functions, we see that the contribution to the amplitude from
the quartic lagrangian is given by
\begin{eqnarray}
\mathcal{L}_{4}&=&-\frac{1}{4}(C^{1234}-S^{1234})(2n D_{22nn}-4n |\vec{x}_{24}|^{2}D_{23nn+1})
\nonumber \\
&+&\frac{1}{8}n(-\delta^{12}_{2}\delta^{34}_{n}+(6+n)C^{1234}+(3n-4)S^{1234}-n\Upsilon^{1234})D_{22nn}
\end{eqnarray}
where again an overall factor of $C(n)^{2}C(2)^{2}$ was omitted, but should be included. We can rewrite this
expression in terms of the $\bar{D}$-functions
\begin{eqnarray}
\mathcal{L}_{4}&=&\pi^2\frac{(C(2)C(n))^{2}}{\Gamma(n)}\frac{u^2}{|\vec{x}_{12}|^{4}|\vec{x}_{34}|^{2n}}\left[-\frac{n}{4}(C^{1234}-S^{1234})(
\bar{D}_{22nn}-\bar{D}_{23nn+1})\right.
\nonumber \\
&+&\left.\frac{1}{2^4}n(-\delta^{12}_{2}\delta^{34}_{n}+(6+n)C^{1234}+(3n-4)S^{1234}-n\Upsilon^{1234})\bar{D}_{22nn}\right]
\label{quarticfinal}
\end{eqnarray}
The final result for the on-shell action is then given by substituting the expressions for the exchange amplitudes on equations (\ref{s-channel}) and (\ref{t-channel}) and by equation (\ref{quarticfinal}).
\subsection{Results for the Four-Point Function}
We collect the results for the relevant on-shell action. First we write down the part of the lagrangian that contributes to the four-point function of interest
\begin{eqnarray}
\mathcal{L}_{on-shell}&=&-n(n-1)\langle C_{2}^{1}C_{2}^{2}C_{2}^{5}\rangle \langle C_{n}^{3}C_{n}^{4}C_{2}^{5}\rangle
\int [dz] s^{1}_{2}(z)s^{2}_{2}(z)G(z,w)s^{3}_{2}(w)s^{4}_{2}(w) 
\nonumber \\
&-&n^2(n-1)^2\langle C_{2}^{1}C_{n}^{3}C_{n}^{5}\rangle \langle C_{2}^{2}C_{n}^{4}C_{n}^{5}\rangle
\int [dz] s^{1}_{2}(z)s^{3}_{n}(z)G(z,w)s^{2}_{2}(w)s^{4}_{n}(w) 
\nonumber \\
&-&\frac{n}{2^4}\langle C^{1}_{2}C^{2}_{2}C^{5}_{[1,0,1]} \rangle \langle C^{3}_{n}C^{4}_{n}C^{5}_{[1,0,1]} \rangle
\int [dz]s^{1}_{2}(z)\buildrel\leftrightarrow\over\nabla^{\mu}s^{2}_{2}(z)
G_{\mu\nu}(z,w)s_{n}^{3}(w)\buildrel\leftrightarrow\over\nabla^{\nu }s^{4}_{n}(w) 
\nonumber \\ 
&-&\frac{n(n-1)}{2^3}\langle C^{1}_{2}C^{3}_{n}C^{5}_{[1,n-1,1]} \rangle \langle C^{2}_{2}C^{4}_{n}C^{5}_{[1,n-1,1]} \rangle
\int [dz]s_{2}^{1}(z)\buildrel\leftrightarrow\over\nabla^{\mu}s_{n}^{3}(z)
G_{\mu\nu}(z,w)s^{2}_{2}(w)\buildrel\leftrightarrow\over\nabla^{\nu}s_{n}^{4}(w) 
\nonumber
\end{eqnarray}
\begin{eqnarray}
&-&\frac{1}{2^4} \langle C^{1}_{2}C^{2}_{2}C^{5}_{[0,0,0]} \rangle \langle C^{3}_{3}C^{4}_{3}C^{5}_{[0,0,0]} \rangle
\int [dz] T^{\mu\nu}_{22}(z)G_{\mu\nu\mu'\nu'}(z,w)T^{\mu'\nu'}_{nn}(w) 
\nonumber \\
&-&\frac{1}{2^3}\langle C^{1}_{2}C^{3}_{n}C^{5}_{[0,n-2,0]} \rangle \langle C^{2}_{2}C^{4}_{n}C^{5}_{[0,n-2,0]} \rangle
\int [dz] T^{\mu\nu}_{2n}(z)G_{\mu\nu\mu'\nu'}(z,w)T^{\mu'\nu'}_{2n}(w) 
\nonumber \\
&-&\frac{1}{2^2}(C^{1234}-S^{1234})s_{2}^{1}(w)\nabla_{\mu}s_{2}^{2}(w)s_{n}^{3}(w)\nabla^{\mu}s_{n}^{4}(w)
\nonumber \\
&+&\frac{1}{2^3}n\left(-\delta^{12}_{2}\delta^{34}_{n}+(6+n)C^{1234}+(3n-4)S^{1234}-n\Upsilon^{1234}\right)s_{2}^{1}(w)s_{2}^{1}(w)s_{n}^{3}(w)s_{n}^{4}(w)
\end{eqnarray}
We now substitute the summation of overlapping $SO(6)$ tensors (see appendix \ref{sec:SphereInts}) and use the results for the exchange integrals. After relabelling the indices, one finally gets the on-shell value of the action that determines the four-point function 
\begin{eqnarray}
\mathcal{S}&=&\frac{N^{2}}{8\pi^{2}}\frac{(n-1)^2(n-2)^2}{4\pi^{6}\Gamma(n)}
\int d^{4}\vec{x}_{1}d^{4}\vec{x}_{2}d^{4}\vec{x}_{3}d^{4}\vec{x}_{4}
s_{2}^{1}(\vec{x}_{1})s_{2}^{2}(\vec{x}_{2})s_{n}^{3}(\vec{x}_{3})s_{n}^{4}(\vec{x}_{4})
\frac{u}{|\vec{x}_{12}|^{4}|\vec{x}_{34}|^{2n}}\left\{\right.
\nonumber \\
&+&\delta^{12}_{2}\delta^{34}_{n}\frac{n}{2^{5}}\left[\bar{D}_{11nn}-(n+1)u\bar{D}_{22nn}-(1+v-u)\bar{D}_{22n+1n+1}\right]
\nonumber \\
&+&C^{1234}\frac{n}{2^4}\left[-2\bar{D}_{11nn}-2(n-1)u\bar{D}_{12n-1n}+(n+6)u\bar{D}_{22nn}-2\bar{D}_{21nn+1}+2\bar{D}_{12nn+1}\right.
\nonumber \\
&-&\left.(u\bar{D}_{31nn}-(n-1)u^2\bar{D}_{23n-1n})-u((n-1)\bar{D}_{22n-1n+1}-\bar{D}_{12nn+1})\right]
\nonumber \\
&+&C^{1243}\frac{n}{2^2}\left[u\bar{D}_{23nn+1}-u\bar{D}_{22nn}\right]
\nonumber  \\
&+&\Upsilon^{1234}\frac{n}{2^4(n+2)}\left[\frac{2(n-1)^{2}(n+2)}{(n+1)}u\bar{D}_{12n-1n}+(n-2)(u\bar{D}_{31nn}-(n-1)u^2\bar{D}_{23n-1n})\right.
\nonumber \\
&+&(n-2)u((n-1)\bar{D}_{22n-1n+1}-\bar{D}_{12nn+1})-n(n+2)u\bar{D}_{22nn}
\nonumber \\
&+&\left.\frac{2}{n+1}(n(n-1)u^2\bar{D}_{33n-1n+1}+2u\bar{D}_{31n+1n+1}+2n(1-u-v)u\bar{D}_{23nn+1})\right]
\nonumber 
\end{eqnarray}
\begin{eqnarray}
&+&S^{1234}\frac{n}{2^4}\left[-2(n-1)^2u\bar{D}_{12n-1n}+3nu\bar{D}_{22nn}
-4u\bar{D}_{23nn+1}\right.
\nonumber \\
&+&\left.\left.(u\bar{D}_{31nn}-(n-1)u^2\bar{D}_{23n-1n})+((n-1)u\bar{D}_{22n-1n+1}-u\bar{D}_{12nn+1})\right]
\right\}
\end{eqnarray}
Here we have made use of some identities relating $\bar{D}$-functions (appendix \ref{sec:Dfunc}) to simplify the expressions. Notice that here we are abusing of the notation, as the scalar fields now refer to the boundary sources, and so depend on the $\vec{x}_i$ coordinates. We are now ready to compute the four-point function (\ref{diffweightprocess}) using the AdS/CFT prescription given in (\ref{prescriptionadscft}). Of course, we need first to canonically normalise the corresponding 1/2-BPS operators, taking into account the rescaling we did to the action at the beginning of this computation
\begin{equation}
\tilde{s}^{I}_{n}=\frac{N}{4\pi^{2}}(n-2)^{1/2}(n-1)s^{I}_{n}
\qquad \qquad
\tilde{s}_{2}^{I}=\frac{N}{4\sqrt{2}\pi^2}s^{I}_{2}
\end{equation}
This implies that the connected piece of the four-point function is of order $\mathcal{O}(1/N^2)$. The explicit form can be determined from
\begin{equation}
\langle \mathcal{O}_{2}(\vec{x}_{1})\mathcal{O}_{2}(\vec{x}_{2})\mathcal{O}_{n}(\vec{x}_{3})\mathcal{O}_{n}(\vec{x}_{4})\rangle=
\frac{2^{9}\pi^{8}}{N^4}\frac{1}{(n-2)(n-1)^2}
\frac{\delta}{\delta s_{2}^{I_{1}}(\vec{x}_{1})}\frac{\delta}{\delta s_{2}^{I_{2}}(\vec{x}_{2})}
\frac{\delta}{\delta s_{n}^{I_{3}}(\vec{x}_{3})}\frac{\delta}{\delta s_{n}^{I_{4}}(\vec{x}_{4})}(-S)
\end{equation}
Upon functional differentiation, the contribution to the amplitude from each of the tensor structures will be given by the corresponding orbit, this is, the $s$, $t$ and $u$ channels obtained by independent permutations of the points $1 \leftrightarrow 2$, $3 \leftrightarrow 4$. Here we make use of the symmetries of the $SO(6)$ tensors, so the final result reads as follows
\begin{eqnarray}
&&\langle \mathcal{O}_{2}^{I_{1}}(\vec{x}_{1})\mathcal{O}_{2}^{I_{2}}(\vec{x}_{2})\mathcal{O}_{n}^{I_{3}}(\vec{x}_{3})\mathcal{O}_{n}^{I_{4}}(\vec{x}_{4})\rangle
=\frac{1}{{\vec{x}_{12}}^4{\vec{x}_{34}^{2n}}}\left\{A(u,v)\delta^{I_1I_2}_{2}\delta^{I_3I_4}_{n}+B_{1}(u,v)C^{I_{1}I_{2}I_{3}I_{4}}
\right.
\nonumber \\
&&\left.+B_{2}(u,v)C^{I_{1}I_{2}I_{4}I_{3}}+c_{1}(u,v)\Upsilon^{I_{1}I_{2}I_{3}I_{4}}+C_{2}(u,v)\Upsilon^{I_{1}I_{2}I_{4}I_{3}}
+D(u,v)S^{I_{1}I_{2}I_{3}I_{4}}\right\}
\label{resultdiffweightsugra}
\end{eqnarray}
where the functions $(A, B_{1}, B_{2}, C_{1}, C_{2}, D)$ are given by 
\begin{equation}
(A, B_{1}, B_{2}, C_{1}, C_{2}, D)=\frac{2^{4}(n-2)}{\Gamma(n)}\frac{1}{N^2}(\tilde{A},\tilde{B}_{1},\tilde{B}_{2},\tilde{C}_1,\tilde{C}_2, \tilde{D})
\end{equation}
and
\begin{eqnarray}
\tilde{A}(u,v)&=&
-\frac{n}{2^3}u\left\{\bar{D}_{11nn}-(n+1)u\bar{D}_{22nn}-(1+v-u)\bar{D}_{22n+1n+1}\right\}
\nonumber \\
\tilde{B}_1(u,v)&=&
-\frac{n}{2^3}u \left\{-2\bar{D}_{11nn}-2(n-1)u\bar{D}_{12n-1n}
-2(\bar{D}_{21nn+1}-\bar{D}_{21n+1n})+(n+6)u\bar{D}_{22nn}\right.
\nonumber \\
&-&(u\bar{D}_{31nn}-(n-1)u^2\bar{D}_{23n-1n})-((n-1)u\bar{D}_{22n-1n+1}-u\bar{D}_{12nn+1})
\nonumber \\
&-&\left.4u(\bar{D}_{22nn}-\bar{D}_{32nn+1})\right\}
\nonumber \\
\tilde{B}_2(u,v)&=&
-\frac{n}{2^3}u\left\{-2\bar{D}_{11nn}-2(n-1)u\bar{D}_{12nn-1}
-2(\bar{D}_{12nn+1}-\bar{D}_{21nn+1})+(n+6)u\bar{D}_{22nn}\right.
\nonumber \\
&-&(u\bar{D}_{13nn}-(n-1)u^2\bar{D}_{23nn-1})-((n-1)u\bar{D}_{22n+1n-1}-u\bar{D}_{12n+1n})
\nonumber \\
&-&\left.4u(\bar{D}_{22nn}-\bar{D}_{23nn+1})\right\}
\nonumber \\
\tilde{C}_1(u,v)&=&
-\frac{n}{2^3(n+2)}u^2\left\{\frac{2(n-1)^2(n+2)}{(n+1)}\bar{D}_{12n-1n}
-n(n+2)\bar{D}_{22nn} \right.
\nonumber \\
&+&\frac{2n(n-1)}{n+1}u\bar{D}_{33n-1n+1}
+\frac{4}{n+1}\bar{D}_{31n+1n+1}+\frac{4n}{n+1}(1-u-v)\bar{D}_{23nn+1}
\nonumber \\
&+&\left.(n-2)((n-1)\bar{D}_{22n-1n+1}-\bar{D}_{12nn+1})+(n-2)(\bar{D}_{31nn}-(n-1)u\bar{D}_{23n-1n})\right\}
\nonumber \\
\tilde{C}_2(u,v)&=&
-\frac{n}{2^3(n+2)}u^2\left\{\frac{2(n-1)^2(n+2)}{(n+1)}\bar{D}_{12nn-1}
-n(n+2)\bar{D}_{22nn}\right.
\nonumber \\
&+&\frac{2n(n-1)}{n+1}u\bar{D}_{33n+1n-1}
+\frac{4}{n+1}\bar{D}_{13n+1n+1}+\frac{4n}{n+1}(v-u-1)\bar{D}_{23n+1n}
\nonumber \\
&+&\left.(n-2)((n-1)\bar{D}_{22n+1n-1}-\bar{D}_{12n+1n})+(n-2)(\bar{D}_{13nn}-
(n-1)u\bar{D}_{23nn-1}\right\}
\nonumber \\
\tilde{D}(u,v)&=&
-\frac{n}{2^3}u^2\left\{-2(n-1)^2(\bar{D}_{12n-1n}+\bar{D}_{12nn-1})+6n\bar{D}_{22nn}-4(\bar{D}_{23nn+1}+\bar{D}_{32nn+1})\right.
\nonumber \\
&+&(n-1)(\bar{D}_{22n-1n+1}+\bar{D}_{22n+1n-1})-(\bar{D}_{12nn+1}+\bar{D}_{12n+1n})
\nonumber \\
&+&\left.(\bar{D}_{31nn}+\bar{D}_{13nn})-(n-1)u(\bar{D}_{23n-1n}+\bar{D}_{23nn-1})\right\}
\label{orbits}
\end{eqnarray}
From (\ref{orbits}) it is possible to see that the crossing symmetries are respected and that the overall form of the four-point amplitude is consistent with conformal symmetry.
\section{Verifying the CFT Predictions}
\label{sec:verifying}
We now try to verify the dynamical constraints imposed on the amplitude by the insertion procedure, on the supergravity result. To do this, we need to rewrite the result (\ref{resultdiffweightsugra}) in a simpler way. We will follow the notation in \cite{Rayson:2007th}, which is based on ideas developed in \cite{Nirschl:2004pa, Dolan:2003hv} and introduce the conformal invariants
\begin{equation}
\sigma=\frac{u_1\cdot u_3 u_2\cdot u_4}{u_1\cdot u_2 u_3 \cdot u_4} \qquad \qquad \tau=\frac{u_1 \cdot u_4 u_2 \cdot u_3}{u_1 \cdot u_2 u_3 \cdot u_4} 
\label{ucrossrad1}
\end{equation}
so the four-point function  (\ref{diffweightprocess}) is given by
\begin{equation}
 \langle \mathcal{O}_{2}(\vec{x}_1,u_1)\mathcal{O}_{2}(\vec{x}_2,u_2)\mathcal{O}_{n}(\vec{x}_3,u_3)\mathcal{O}_{n}(\vec{x}_4,u_4)\rangle =\left(\frac{u_1.u_2}{|\vec{x}_{12}|^2}\right)^{2}\left(\frac{u_3.u_4}{|\vec{x}_{34}|^2}\right)^{n}\mathcal{G}^{(2,2,n,n)}(u,v;\sigma,\tau)
\end{equation}
where
\begin{equation}
\mathcal{G}^{(2,2,n,n)}(u,v;\sigma,\tau)=\mathcal{G}_{0}(u,v;\sigma,\tau)+s(u,v;\sigma,\tau)\mathcal{H}_{I}(u,v;\sigma,\tau)
\label{generalstrucG}
\end{equation}
$\mathcal{H}_I$ contains all the non-trivial dynamic contributions and $\mathcal{G}_{0}$ is the free field part, which has the following structure
\begin{equation}
\mathcal{G}_{0}(u,v;\sigma,\tau)=k+G_{f}(u,v;\sigma,\tau)+s(u,v;\sigma,\tau)\mathcal{H}_{0}(u,v,\sigma,\tau)
\end{equation}
In these expressions
\begin{equation}
s(u,v;\sigma,\tau)=v+\sigma^2 uv +\tau^2 u +\sigma v (v-1-u)+\tau (1-u-v) + \sigma \tau u (u-1-v)
\end{equation}
The free field term in the $22\rightarrow nn$ channel is given by the expression \cite{Rayson:2007th,Nirschl:2004pa, Dolan:2003hv}
\begin{equation}
\label{freeg}
\mathcal{G}_0(u,v;\sigma,\tau)=1+b_1\left(\sigma u  +\tau \frac{u}{v}\right)+c_1\left(\sigma^2 u^2 +\tau^2 \frac{u^2}{v^2}\right)+d \sigma \tau \frac{u}{v}
\end{equation}
with $b_1, c_1$ and $d$ are given in (\ref{largeNfreecoeff}). The $2n \rightarrow 2n$ channel can be obtained using crossing symmetry. From (\ref{resultdiffweightsugra}), one can read the expression in the interacting theory 
\begin{equation}
\mathcal{G}(u,v;\sigma,\tau)=a(u,v)+\left(\sigma u b_{1}(u,v)+\tau\frac{u}{v}b_{2}(u,v) \right)+\left(\sigma^2 u^2 c_{1}(u,v)+\tau^2\frac{u^2}{v^2}c_{2}(u,v)\right)+\sigma\tau \frac{u^2}{v}d(u,v)
\label{intg}
\end{equation}
where $a(u,v)=A(u,v)$, $b_1(u,v)=\frac{B_{1}(u,v)}{u}$, $c_1(u,v)=\frac{C_{1}(u,v)}{u^2}$ and $d(u,v)=\frac{v}{u^2}D(u,v)$. $b_2(u,v)$ and $c_2(u,v)$ can be obtained from crossing symmetry, as the supergravity result (\ref{largeNfreecoeff}) satisfies this property. Notice also that the cross-ratios $\sigma$ and $\tau$ defined in (\ref{ucrossrad1}) arise naturally from expressing the products of $C$-tensors in terms of harmonic polynomials (see appendix \ref{sec:HarmPoly}).

It is possible to rewrite (\ref{intg}) by simplifying the result (\ref{largeNfreecoeff}), using identities between $\bar{D}$-functions (see appendix \ref{sec:Dfunc}). The simplification was done in \cite{Rayson:2007th} and we reproduce it here. One gets
\begin{equation}
\mathcal{G}(u,v;\sigma,\tau)=1+\frac{2n}{N^2}\left(\sigma u +\tau \frac{u}{v}+(n-1)\sigma\tau \frac{u^2}{v}-\frac{1}{(n-2)!}s(u,v;\sigma,\tau)u^n\bar{D}_{nn+222}(u,v) \right)
\end{equation}
where the disconnected piece has been normalised to 1. In the free field limit, $\mathcal{G}\rightarrow \mathcal{G}_0$, so comparing (\ref{freeg}) with (\ref{intg}) one has
\begin{equation}
a(u,v)\rightarrow 1 \qquad b_i(u,v) \rightarrow b_i \qquad c_i(u,v) \rightarrow c_i \qquad d(u,v) \rightarrow d
\end{equation}
from where we can identify $k=1+(n+1)b_i+2c_i$ and from (\ref{generalstrucG}) one sees that\footnote{This can be read of from $a(u,v)$ as its connected piece has no free field contributions.}
\begin{equation}
\mathcal{H}_I(u,v)=-\frac{2n}{N^2}\frac{1}{(n-2)!}u^n \bar{D}_{nn+222}(u,v)
\label{Hint}
\end{equation}
In the $2n\rightarrow 2n$ channel the previous expression reads
\begin{equation}
\hat{\mathcal{H}}_I(u,v)=-\frac{2n}{N^2}\frac{1}{(n-2)!}u^2 \bar{D}_{2n+22n}(u,v)
\end{equation}
It is now clear that one can write
\begin{eqnarray}
a(u,v)&=&1+v \mathcal{H}_I(u,v) \hspace{28.mm} d(u,v)=d+\frac{v}{u}(u-v-1)\mathcal{H}_I(u,v)
\nonumber\\
b_{1}(u,v)&=&b_1+\frac{v}{u}(v-u-1)\mathcal{H}_I(u,v) \qquad b_{2}(u,v)=b_2+\frac{v}{u}(1-u-v)\mathcal{H}_I(u,v)
\nonumber \\
c_{1}(u,v)&=&c_1+\frac{v}{u}\mathcal{H}_I(u,v)  \hspace{25.mm} c_{2}(u,v)=c_2+\frac{v^2}{u}\mathcal{H}_I(u,v)
\label{splitcoeff}
\end{eqnarray}
so the supergravity result also splits into a free and a quantum part, as it was predicted by superconformal symmetry. Defining 
\begin{equation}
\mathcal{H}_I(u,v)=\frac{u}{v}\mathcal{F}(u,v)
\end{equation}
it becomes clear that the relations (\ref{insertrels}) are satisfied. We consider this fact as a strong evidence in favour of the AdS/CFT correspondence. 

We can also read off the values of the coefficients $b_i, c_i$ and $d$ from the free part of the function $\mathcal{G}(u,v;\sigma,\tau)$. The results are
\begin{equation}
b_i=\frac{2n}{N^2} \qquad c_i=0 \qquad d = \frac{2n(n-1)}{N^2} 
\label{cizero}
\end{equation}
Notice that the values of $b_i$ and $d$ agree with those computed using free field theory. This is a highly non-trivial result. However, $c_i$ vanish, which is apparently at odds with what was obtained using free fields, but recall that $c_i$ was dependent on the colour structure of the operators. This might suggest that this quantity receives quantum corrections. It should also be noticed that in the case $n=3$, one has $c_i=0$ so there is agreement \cite{Berdichevsky:2007xd}. 

\section{Conclusions and Outlook}
\label{sec:conclusions}
In this paper, we have investigated four-point functions of different weight operators in the context of the AdS/CFT correspondence. We have looked at a specific computations in the supergravity approximation (large $\lambda$, large $N$), of a process involving fields dual to primaries of conformal dimension $2$ and primaries of conformal dimension $n$. The results have been analysed using results from free field Yang-Mills theory and superconformal symmetry. Some of our key results are summarised below:
\begin{itemize}
\item The connected piece of the four-point function of 1/2-BPS superconformal primaries of conformal weights 2 and $n$, was shown to have a structure that is consistent with superconformal symmetry. Moreover, we have seen it naturally separates into a free and an interacting (quantum) piece, which involves all the non-trivial dynamics and satisfies the restrictions imposed by the insertion procedure. 

\item A new method was used for evaluating effective couplings in the lagrangian arising from integrals over $S^5$. This allowed the determination of the on-shell lagrangian for KK scalars dual to superconformal primaries in the YM side. 

\item We provided further evidence for the possibility that the quartic four-derivative Lagrangian of \cite{Arutyunov:1999fb} vanishes, as now we have extended the computation of the lagrangian to include primaries with different conformal weights, with two of them being generic (i.e. no specification of the representation content). As it has been argued before in \cite{Arutyunov:2002fh, Arutyunov:2003ae}, this would imply the existence of a $\sigma$-model action describing the extension of  $d=5$ $\mathcal{N}=8$ supergravity to include massive KK modes of the IIB compactification.
\end{itemize}
With the techniques developed in appendix \ref{sec:HarmPoly} to compute the interaction couplings arising from the products of $C$-tensors, it seems likely that the computation of the correlation function 
\begin{equation*}
\langle \mathcal{O}_{n_1}(\vec{x}_1)\mathcal{O}_{n_1}(\vec{x}_2)\mathcal{O}_{n_2}(\vec{x}_3)\mathcal{O}_{n_2}(\vec{x}_4)\rangle
\end{equation*} 
in AdS supergravity could be evaluated.  This would give us further information on the dynamics of KK scalars, and would provide additional evidence for the vanishing of the quartic four-derivative lagrangian in the five-dimensional effective theory.

Another problem one could explore is the effect of $\mathcal{R}^4$ corrections to four-point functions of superconformal primary operators. Recalling that the dual fields are built from the trace of the graviton in the $S^5$ and the RR four-form on $S^5$ and given that all the terms at order $\alpha'^3$ involving the metric and the four-form are known from \cite{Paulos:2008tn}, it is conceivable that the corrections to the five-dimensional effective lagrangian can be obtained. This indeed would be a difficult task, but a first step would be to consider the case of lowest scale dimension primaries($\Delta=2$). In this way, it should be possible to compute the order $(g_{YM}^2N)^{-3/2}$ correction to the four-point function of lowest weight primaries.  

A puzzle that remains to be addressed is the mismatch of the $c_i$ coefficient function from the supergravity computation, eq. (\ref{cizero}), and the free-field theory one,  eq. (\ref{largeNfreecoeff}). Given that the supergravity result gives $c_i=0$, one might imagine that there should be stringy corrections to this quantity. Corrections in $\alpha'$ could be considered once the higher order corrections to the five-dimensional effective action are known. Another interesting avenue would be to consider the potential contribution coming from non-perturbative effects \cite{Green:2002vf}.

Finally it should be mentioned that the supergravity result obtained here can be used to ana\-lyse the structure of the OPE of the primaries at strong coupling and to evaluate anomalous dimensions. Some results in this matter can be found in \cite{Rayson:2007th}.
\\
\appendix
\section{Integrals over the Sphere}
\label{sec:SphereInts}
Upon reduction of the ten-dimensional action, the supergravity fields couple through $SO(6)$ invariant tensors which are given by integrals of spherical harmonics on the five-sphere
\begin{equation}
a_{123}=\int Y^{I_{1}}Y^{I_{2}}Y^{I_{3}} \qquad t_{123}=\int \nabla^{\alpha}Y^{I_{1}}Y^{I_{2}}Y^{I_{3}}_{\alpha}
\qquad p_{123}=\int \nabla^{\alpha}Y^{I_{1}}\nabla^{\beta}Y^{I_{2}}Y^{I_{3}}_{(\alpha \beta)}
\end{equation}
All irreducible representations of $SO(6)$ that are required, can always be expressed in terms of canonically normalized $C$-tensors with corresponding Young symmetry. The integrals of spherical harmonics can then be expressed in terms of $C$-tensors as follows
\begin{eqnarray}
a_{123}&=&\frac{\prod_{i=1}^{3}\frac{k_{i}!z(k_{i})}{\alpha_{i}!}}{\pi^{3/2}(\sigma+2)!2^{\sigma-1}}
\langle C_{[0,k_{1},0]}^{1}C_{[0,k_{2},0]}^{2}C_{[0,k_{3},0]}^{3}\rangle \nonumber \\
t_{123}&=&\frac{\prod_{i=1}^{3}\frac{k_{i}!z(k_{i})}{(\alpha_{i}-\frac{1}{2})!}}{\pi^{3/2}(k_{3}+1)(\sigma+\frac{3}{2})!2^{\sigma-\frac{3}{2}}}
\langle C_{[0,k_{1},0]}^{1}C_{[0,k_{2},0]}^{2}C_{[1,k_{3}-1,1]}^{3}\rangle \nonumber \\
p_{123}&=&\frac{\alpha_{3}\prod_{i=1}^{3}\frac{k_{i}!z(k_{i})}{\alpha_{i}!}}{\pi^{3/2}(\sigma+1)!2^{\sigma}}
\langle C_{[0,k_{1},0]}^{1}C_{[0,k_{2},0]}^{2}C_{[2,k_{3}-2,2]}^{3}\rangle
\end{eqnarray}
where $z(k)=(2^{k-1}(k+1)(k+2))^{1/2}$, $\sigma=\frac{1}{2}(k_{1}+k_{2}+k_{3})$ and $\alpha_{i}=\frac{1}{2}(k_{j}+k_{l}-k_{i})$.
Here the notation we follow stands for
\begin{equation}
\langle C_{[0,k_{1},0]}^{1}C_{[0,k_{2},0]}^{2}C_{[0,k_{3},0]}^{3}\rangle=C^{I_{1}}_{i_{1}...i_{\alpha_{2}}j_{1}...j_{\alpha_{3}}}
C^{I_{2}}_{j_{1}...j_{\alpha_{3}}l_{1}...l_{\alpha_{1}}}C^{I_{3}}_{l_{1}...l_{\alpha_{1}}i_{1}...i_{\alpha_{2}}}
\end{equation}
and
\begin{eqnarray}
\langle C_{[0,k_{1},0]}^{1}C_{[0,k_{2},0]}^{2}C_{[1,k_{3}-1,1]}^{3}\rangle&=&C^{I_{1}}_{mi_{1}...i_{p_{2}}j_{1}...j_{p_{3}}}
C^{I_{2}}_{j_{1}...j_{p_{3}}l_{1}...l_{p_{1}}}C^{I_{3}}_{m;l_{1}...l_{p_{1}}i_{1}...i_{p_{2}}}
\nonumber \\
&-&C^{I_{1}}_{i_{1}...i_{p_{2}+1}j_{1}...j_{p_{3}}}C^{I_{2}}_{j_{1}...j_{p_{3}}l_{1}...l_{p_{1}-1}}
C^{I_{3}}_{m;l_{1}...l_{p_{1}-1}i_{1}...i_{p_{2}+1}}
\end{eqnarray}
where $p_{1}=\alpha_{1}+\frac{1}{2}$, $p_{2}=\alpha_{2}-\frac{1}{2}$ and $p_{3}=\alpha_{3}-\frac{1}{2}$. Finally,
\begin{equation}
\langle C_{[0,k_{1},0]}^{1}C_{[0,k_{2},0]}^{2}C_{[2,k_{3}-2,2]}^{3}\rangle=C^{I_{1}}_{mi_{1}...i_{p_{2}}j_{1}...j_{p_{3}}}
C^{I_{2}}_{nj_{1}...j_{p_{3}}l_{1}...l_{p_{1}}}C^{I_{3}}_{mn;l_{1}...l_{p_{1}}i_{1}...i_{p_{2}}}
\end{equation}
From the AdS exchange diagrams and the quartic couplings, we see that one needs to express products of the form $\langle C^{1}C^{2}C^{5}\rangle \langle C^{3}C^{4}C^{5}\rangle$, where summation over the representation of the fifth index is assumed, in terms of a basis of independent tensor structures. The product can be expressed in terms of combinations of Kronecker deltas \cite{Arutyunov:2002fh}. One has
\begin{equation}
C^{I}_{i_{1}...i_{n}}C^{I}_{j_{1}...j_{n}}=\sum_{k=0}^{\left[\frac{n}{2}\right]}\theta_{k}\sum_{(l_{2k-1}...l_{2k})}
\delta_{i_{l_{1}}i_{l_{2}}}...\delta_{i_{l_{1}}i_{l_{2}}}...\delta_{i_{l_{2k-1}}i_{l_{2k}}}
\delta^{(n-2k)}_{i_{1}...\hat{i}_{l_{1}}...\hat{i}_{l_{2k}}...i_{l_{n}},(j_{2k+1}...j_{n}}\delta_{j_{1}j_{2}}...\delta_{j_{2k-1}j_{2k})}
\label{completeness}
\end{equation}
where $(...)$ stands for total symmetrisation of indices and $\delta^{(p)}_{i_{1}...i_{p},j_{1}...j_{p}}=\delta^{(p)}_{(i_{1}...i_{p}), (j_{1}...j_{p})}$ denotes the symmetrised product of $p$ kronecker deltas $\delta_{i_{r}j_{s}}$. The coefficients $\theta_{k}$ are given by
\begin{equation}
\theta_{0}=1 \qquad \theta_{k}=\frac{(-1)^{k}}{2^{k}(n+1)...(n+2-k)}
\end{equation}
Evidently (\ref{completeness}) it is useful when one is dealing with correlation functions involving chiral primaries of lower weight. However, its application becomes increasingly involved once one has higher rank tensors.  One needs then to develop some other method to determine the sums of products of $SO(6)$ tensors, that enter the amplitude.
\section{Harmonic Polynomials}
\label{sec:HarmPoly}
We reproduce here some results derived in \cite{Dolan:2003hv,Dolan:2006ec} \footnote{We thank H. Osborn for bringing these results to our attention and suggesting the use of harmonic polynomials to obtain the couplings.}. One needs to consider the expansion of four point functions in terms of the eigenfunctions of the $SO(6)$ Casimir operator
\begin{equation}
L^{2}=\frac{1}{2}L_{ab}L_{ab}
\end{equation}
where the generators are given by
\begin{equation}
L_{ab}=u_{1a}\partial_{1b}-u_{1b}\partial_{1a}+u_{2a}\partial_{2b}-u_{2b}\partial_{2a}
\end{equation}
which is expressed in terms of null vectors $u_{1}$, $u_{2}$, $u_{3}$, $u_{4}$. One can prove that
\begin{equation}
L_{ab}u_{1}\cdot u_{2}=0
\end{equation}
so that
\begin{equation}
L^{2}(u_{1}\cdot t_{2})^{k}(u_{3}\cdot u_{4})^{l}f(\sigma,\tau)=(u_{1}\cdot u_{2})^{k}(u_{3}\cdot u_{4})^{l}L^{2}f(\sigma,\tau)
\end{equation}
where $\sigma$ and $\tau$ are given by
\begin{equation}
\sigma=\frac{u_{1}\cdot u_{3} u_{2}\cdot u_{4}}{u_{1}\cdot u_{2} u_{3} \cdot u_{4}}
\qquad \qquad
\tau=\frac{u_{1}\cdot u_{4} u_{2}\cdot u_{3}}{u_{1}\cdot u_{2} u_{3} \cdot u_{4}}
\label{ucrossratios}
\end{equation}
so that one can consider eigenfunctions which are polynomials in $\sigma$, $\tau$.
\begin{equation}
Y(\sigma,\tau)=\sum_{t\ge 0}\sum_{q=0}^{t}c_{t,q}\sigma^{t-q}\tau^{q}
\label{harmpoly}
\end{equation}
which satisfies the eigenvalue equation
\begin{equation}
L^{2}Y(\sigma,\tau)=-2CY(\sigma,\tau)
\label{evaluegen}
\end{equation}
If $t_{max}=n$, it is possible to solve for the coefficients in the expansion (\ref{harmpoly}) and for a given $n$, there will be $m=n+1$ eigenfunctions orthogonal with respect to integration over $\sigma,\tau \ge 0$, $\sqrt{\sigma}+\sqrt{\tau} \le 1$.

Up to a normalisation constant, each term may be identified with terms in the projection operators on irreducible representations of $SO(6)$, where $Y_{nm}$ corresponds to the $SU(4)$$\simeq$ $SO(6)$ representation with Dynkin labels $[n-m,2m,n-m]$.

More general forms can be considered when discussing four-point functions in which each field belongs to the same $SO(6)$ representation. For the more general case, one can generalize (\ref{evaluegen}) to
\begin{equation}
L^{2}((u_{1}\cdot u_{4})^{a}(u_{2}\cdot u_{4})^{b}Y^{(a,b)}(\sigma,\tau))=
-2C(((u_{1}\cdot u_{4})^{a}(u_{2}\cdot u_{4})^{b}Y^{(a,b)}(\sigma,\tau))
\end{equation}
In this case, $Y_{nm}^{(a,b)}$ will correspond to the representation $[n-m,a+b+2m,n-m]$. Proceeding accordingly, one can built the lowest eigenfunctions by hand. The ones that are needed are listed below
\begin{eqnarray}
Y_{00}^{(a,0)}&=&1 \nonumber \\
Y_{10}^{(a,0)}&=&\Big(\sigma-\tau+\frac{a}{a+4}\Big) \nonumber \\
Y_{11}^{(a,0)}&=&\Big(\sigma+\frac{\tau}{a+1}-\frac{1}{a+3}\Big) \nonumber \\
Y_{20}^{(a,0)}&=&\Big(\sigma^{2}+\tau^{2}-2\sigma\tau+\frac{a-3}{a+6}\sigma-\frac{2a+3}{a+6}\tau
+\frac{a^{2}+2a+3}{(a+5)(a+6)}\Big) \nonumber \\
Y_{21}^{(a,0)}&=&\Big(\sigma^{2}-\frac{\tau^2}{a+1}-\frac{a}{a+1}\sigma\tau+\frac{a-3}{a+6}\sigma
\nonumber \\
&+&\frac{2a+3}{(a+1)(a+6)}\tau-\frac{a}{(a+4)(a+6)}\Big) \nonumber \\
Y_{22}^{(a,0)}&=&\Big(\sigma^{2}+\frac{2}{(a+1)(a+2)}\tau^{2}+\frac{4}{a+1}\sigma\tau-\frac{4}{a+5}\sigma
\nonumber \\
&-&\frac{4}{(a+1)(a+5)}\tau+\frac{2}{(a+4)(a+5)}\Big)
\end{eqnarray}
The polynomials $Y_{00}^{(0,0)}$, $Y_{11}^{(0,0)}$ and $Y_{22}^{(0,0)}$ give the products of scalar harmonics $a_{125}a_{235}$ for fixed $k_{5}=0,2,4$, in the $s$-channel.  The same polynomials but with $a=n-2$ give the results for the $t$-channel, with $k_{5}=n-2,n-n+2$. The polynomials  $Y_{10}^{(0,0)}$ and $Y_{20}^{(0,0)}$ give the products of vector harmonics $t_{125}t_{345}$ for $k_{5}=1,3$ in the $s$-channel, while in the $t$-channel, $k_{5}=n-1,n+1$, with $a=n-2$. Finally,  the polynomial  $Y_{21}^{(0,0)}$ gives the product of tensor harmonics $p_{125}p_{345}$ for $k_{5}=2$ in the $s$-channel and for $k_{5}=n$, again with $a=n-2$. The results are correct up to an appropriate normalisation constant. By using the completion relation (\ref{completeness}) involving $SO(6)$ tensors, it is possible to fix it so that one can reproduce the results involving $p=2,3$. We first introduce the relation between the monomials in $\sigma$ and $\tau$, with the different tensor structures entering the amplitude, which we list below
\begin{eqnarray}
\delta_{2}^{12}\delta_{n}^{34}&=&C_{ij}^{1}C_{ij}^{2}C_{k_{1}\cdots k_{n}}^{3}C_{k_{1}\cdots k_{n}}^{4} \nonumber \\
C^{1234}&=&C_{ij}^{1}C_{jk}^{2}C_{kl_{1}\cdots l_{n-1}}^{3}C_{il_{1}\cdots l_{n-1}}^{4} \nonumber \\
\Upsilon^{1234}&=&C_{ij}^{1}C_{lm}^{2}C_{ijk_{1}\cdots k_{n-2}}^{3}C_{lmk_{1}\cdots k_{n-2}}^{4} \nonumber \\
S^{1234}&=&C_{ik}^{1}C_{jl}^{2}C_{lk m_{1}\cdots m_{n-2}}^{3}C_{ij m_{1}\cdots m_{n-2}}^{4}
\end{eqnarray}
One then obtains the following formulae. For the $s$-channel
\begin{eqnarray}
\sigma^{2}&\equiv& \Upsilon^{1234} \qquad \tau^{2} \equiv \Upsilon^{1243}  \nonumber \\
\sigma &\equiv& C^{1234} \qquad \tau \equiv C^{1243} \nonumber  \\
\sigma \tau &\equiv& S^{1234} \qquad 1 \equiv \delta^{12}_{2}\delta^{34}_{n}  \nonumber
\end{eqnarray}
and for the $t$-channel
\begin{eqnarray}
\tilde{\sigma}^{2}&\equiv& \Upsilon^{1342} \qquad \tilde{\tau}^{2}\equiv \delta^{13}_{2}\delta^{24}_{n}  \nonumber \\
 \tilde{\sigma} &\equiv& S^{1324} \qquad \tilde{\tau} \equiv C^{1342} \nonumber  \\
\tilde{\sigma} \tilde{\tau} &\equiv& C^{1324} \qquad \tilde{1} \equiv \Upsilon^{1324} \nonumber
\end{eqnarray}
where $S^{1234}$ is symmetric under exchange of $1 \leftrightarrow 2$ and $3 \leftrightarrow 4$, while
$C^{1234}$ and $\Upsilon^{1234}$ obey the relations
\begin{equation}
C^{1234}=C^{2143} \qquad \Upsilon^{1234}=\Upsilon^{2143}
\end{equation}
The tilded variables are related to the original ones by
\begin{equation}
\tilde{\sigma}=\frac{1}{\sigma} \qquad  \qquad  \tilde{\tau}=\frac{\tau}{\sigma}
\end{equation}

We now list the expressions that are required by the computation. For the $s$-channel, we set $k_1=k_2=2, k_3=k_4=n$. The contributions from scalar harmonics yield
\begin{equation}
  \begin{split}
  \inp{C^1_2 C^2_2 C^5_{[0,0,0]}}\inp{C^3_n C^4_n C^5_{[0,0,0]}} &= \delta_2^{12} \delta_n^{34}, \\
  \inp{C^1_2 C^2_2 C^5_{[0,2,0]}}\inp{C^3_n C^4_n C^5_{[0,2,0]}} &= \frac{1}{2} C^{1234} +
  \frac{1}{2} C^{1243} - \frac{1}{6} \delta_{2}^{12} \delta_{n}^{34}, \\
  \inp{C^1_2 C^2_2 C^5_{[0,4,0]}}\inp{C^3_n C^4_n C^5_{[0,4,0]}} &= -\frac{2}{15} C^{1234}
  -\frac{2}{15}C^{1243} + \frac{2}{3} S^{1234} \\
  & \quad +\frac{1}{6} \Upsilon^{1243} + \frac{1}{6} \Upsilon^{1234} +\frac{1}{60} \delta_2^{12} \delta_n^{34}.
\end{split}
\label{eq:asum22nn}
\end{equation}
It is clear that these expressions are identical to those entering previous computations, and in view of the formalism involving harmonic polynomials, it is easy to convince oneself that it has to be true. For the summation over the vector representations one gets
\begin{equation}
  \begin{split}
    \inp{C^1_2 C^2_2 C^5_{[1,0,1]}}\inp{C^3_n C^4_n C^5_{[1,0,1]}} &= 2(C^{1243}-C^{1234}), \\
    \inp{C^1_2 C^2_2 C^5_{[1,2,1]}}\inp{C^3_n C^4_n C^5_{[1,2,1]}} &= \frac{1}{3}(C^{1234}-C^{1243})
    + \frac{2}{3}(\Upsilon^{1234} - \Upsilon^{1243}).
  \end{split}
  \label{eq:tsum22nn}
\end{equation}
And for the tensor representation,
\begin{equation}
  \begin{split}
    \inp{C^1_2 C^2_2 C^5_{[2,0,2]}}\inp{C^3_n C^4_n C^5_{[2,0,2]}} &= -\frac{2}{3}\left(C^{1234} +
    C^{1243}\right) + \frac{4}{3}\left(\Upsilon^{1234} + \Upsilon^{1243}\right) \\
    & \quad -\frac{8}{3} S^{1234} + \frac{2}{15} \delta_2^{12} \delta_n^{34}.
  \end{split}
\end{equation}
Next we consider the $t$-channel case in which we set $k_1 = k_3 = 2$ and $k_2 = k_4 = n$. A priori it is possible to see that the normalisation constants will depend on $n$, and we need to determine these first. We do so, by computing various cases in which $n$ takes fixed values\footnote{For this task, we used Cadabra, which is very well suited for doing tensor computations in particular bases \cite{Peeters:2007wn}.}. For summation over scalar harmonics, one gets
\begin{equation}
  \begin{split}
    \inp{C^1_2 C^2_n C^5_{[0,n-2,0]}}\inp{C^3_2 C^4_n C^5_{[0,n-2,0]}} &= \Upsilon^{1324}, \\
    \inp{C^1_2 C^2_n C^5_{[0,n,0]}}\inp{C^3_2 C^4_n C^5_{[0,n,0]}} &= \frac{n-1}{n}\left[S^{1324} +
    \frac{1}{n-1} C^{1342} - \frac{1}{n+1} \Upsilon^{1324}\right], \\
    \inp{C^1_2 C^2_n C^5_{[0,n+2,0]}}\inp{C^3_2 C^4_n C^5_{[0,n+2,0]}} &= \frac{n(n-1)}{(n+1)(n+2)}\left[\Upsilon^{1342}
    +\frac{2}{n(n-1)} \delta_{2}^{13}\delta_{n}^{24} + \frac{4}{n-1} C^{1324}\right. \\
    & \quad \left.- \frac{4}{n+3} S^{1324} - \frac{4}{(n-1)(n+3)} C^{1342} + \frac{2}{(n+2)(n+3)}\Upsilon^{1324}\right],
  \end{split}
\end{equation}
One the scalar contributions are determined, it is easy to compute  the vector and tensor ones by using the following identities\footnote{We thank L. Berdichevsky for the alternative expression for $d_{125}$.}
\begin{eqnarray}
t_{125}t_{345}&=&-\frac{(f_{1}-f_{2})(f_{3}-f_{4})}{4f_{5}}a_{125}a_{345}+\frac{1}{4}(a_{145}a_{235}-a_{245}a_{135})
\nonumber \\
p_{125}p_{345}&=&-\frac{(f_{1}-f_{2})(f_{3}-f_{4})}{2(f_{5}-5)}t_{125}t_{345}
-\frac{5}{4f_{5}(f_{5}-5)}d_{125}d_{345}
\nonumber \\
&-&\frac{1}{20}(f_{1}+f_{2}-f_{5})(f_{3}+f_{4}-f_{5})a_{125}a_{345}
+\frac{1}{8}(f_{1}+f_{3}-f_{5})(f_{2}+f_{4}-f_{5})a_{135}a_{245}
\nonumber \\
&+&\frac{1}{8}(f_{1}+f_{4}-f_{5})(f_{2}+f_{3}-f_{5})a_{145}a_{235}
\end{eqnarray}
where $f_{k}=k(k+4)$ and
\begin{equation}
d_{125}=\left(\frac{1}{10}f_{2}f_{5} +\frac{1}{10}f_{1}f_{5} + \frac{1}{2}f_{1}f_{2}
-\frac{1}{4}f_{1}^{2}-\frac{1}{4}f_{2}^{2}+\frac{3}{20}f_{5}^{2}\right)a_{125}
\end{equation}
Hence the vector contributions read
\begin{equation}
  \begin{split}
    \inp{C^1_2 C^2_n C^5_{[1,n-2,1]}}\inp{C^3_2 C^4_n C^5_{[1,n-2,1]}} &= -\frac{n}{n-1}\left[ S^{1324} -
    C^{1342} + \frac{n-2}{n+2} \Upsilon^{1324}\right], \\
    \inp{C^1_2 C^2_n C^5_{[1,n,1]}}\inp{C^3_2 C^4_n C^5_{[1,n,1]}} &= -\frac{(n-1)(n+2)}{n(n+1)}\left[ \Upsilon^{1342} -
    \frac{1}{n-1} \delta_{2}^{13}\delta_{n}^{24} - \frac{n-2}{n-1} C^{1324} \right. \\
    & \quad \left.+ \frac{n-5}{n+4} S^{1324} + \frac{2(n-2)+3}{(n-1)(n+4)} C^{1342} -
    \frac{n-2}{(n+2)(n+4)} \Upsilon^{1324}\right].
  \end{split}
  \label{eq:tsum2n2n}
\end{equation}
and finally, the tensor case gives
\begin{equation}
    \begin{split}
      \inp{C^1_2 C^2_n C^5_{[2,n-2,2]}}\inp{C^3_2 C^4_n C^5_{[2,n-2,2]}} &= \frac{16(n-1)}{n^2(n+1)}\left[ \Upsilon^{1342} +
      \delta_{2}^{13}\delta_{n}^{24} - 2 C^{1324}+ \frac{n-5}{n+4} S^{1324} \right. \\
      & \quad \left. - \frac{2(n-2)+3}{n+4} C^{1342} + \frac{(n-2)^{2}+2(n-2)+3}{(n+3)(n+4)}
      \Upsilon^{1324}\right].
    \end{split}
\end{equation}
The results of the remaining cases are the same, if one changes the representation labels accordingly, except for equation~\eqref{eq:tsum2n2n}, which acquires an additional minus sign in the cases $k_1=k_4=n, k_2=k_3=2$ and $k_1=k_4=2, k_2=k_3=n$.
\section{Properties of $D$-Functions}
\label{sec:Dfunc}
We collect here the general properties and identities involving the $D$-functions.  These are defined as integrals over $AdS_{5}$, by the formula
\begin{equation}
D_{\Delta_{1}\Delta_{2}\Delta_{3}\Delta_{4}}(\vec{x}_{1},\vec{x}_{2},\vec{x}_{3},\vec{x}_{4})=\int
\frac{d^{5}z}{z_{0}^{5}}\tilde{K}_{\Delta_{1}}(z,\vec{x}_{1})\tilde{K}_{\Delta_{2}}(z,\vec{x}_{2})
\tilde{K}_{\Delta_{3}}(z,\vec{x}_{3})\tilde{K}_{\Delta_{4}}(z,\vec{x}_{4})
\end{equation}
with
\begin{equation}
\tilde{K}_{\Delta}(z,\vec{x})=\left(\frac{z_{0}}{z_{0}^2+(\vec{z}-\vec{x})^{2}}\right)^{\Delta}
\label{Kpropapp}
\end{equation}
$D$-integrals have also a representation in terms of integrals over Feynman parameters
\begin{equation}
D_{\Delta_{1}\Delta_{2}\Delta_{3}\Delta_{4}}(\vec{x}_{1},\vec{x}_{2},\vec{x}_{3},\vec{x}_{4})
=\frac{\pi^{2}\Gamma(\Sigma-2)\Gamma(\Sigma)}{2\prod_{i}\Gamma(\Delta_{i})}\int
\prod_{j}d\alpha_{j}\alpha_{j}^{\Delta_{j}-1}\frac{\delta(\sum_{j}\alpha_{j}-1)}{(\sum_{k<l}\alpha_{k}
\alpha_{l}x_{kl}^2)^{\Sigma}}
\label{Dfeynmanp}
\end{equation}
where $2\Sigma=\sum_{i}\Delta_{i}$. Immediately one can see that any $D$-function can be obtained by differentiation of the box-integral:
\begin{equation}
B(x_{ij})=\int
\prod_{j}d\alpha_{j}\frac{\delta(\sum_{j}\alpha_{j}-1)}{(\sum_{k<l}\alpha_{k}\alpha_{l}x_{kl}^2)^{\Sigma}}
\end{equation}
We define now the $\bar{D}$-functions, which are functions of
conformal invariant ratios, $u$ and $v$, by
\begin{equation}
\bar{D}_{\Delta_{1}\Delta_{2}\Delta_{3}\Delta_{4}}(u,v)=\kappa
\frac{|\vec{x}_{31}|^{2\Sigma-2\Delta_{4}}|\vec{x}_{24}|^{2\Delta_{2}}}
{|\vec{x}_{41}|^{2\Sigma-2\Delta_{1}-2\Delta_{4}}|\vec{x}_{34}|^{2\Sigma-2\Delta_{3}-2\Delta_{4}}}
D_{\Delta_{1}\Delta_{2}\Delta_{3}\Delta_{4}}
\label{Dbardef}
\end{equation}
where
\begin{equation}
\kappa=\frac{2}{\pi^2}\frac{\Gamma(\Delta_{1})\Gamma(\Delta_{2})\Gamma(\Delta_{3})\Gamma(\Delta_{4})}{\Gamma(\Sigma-2)}
\end{equation}
One can obtain identities relating different $\bar{D}$-functions by using the differentiation. These are
\begin{eqnarray}
\bar{D}_{\Delta_{1}+1\Delta_{2}+1\Delta_{3}\Delta_{4}}&=&-\partial_{u}\bar{D}_{\Delta_{1}\Delta_{2}\Delta_{3}\Delta_{4}}
\nonumber \\
\bar{D}_{\Delta_{1}\Delta_{2}+1\Delta_{3}+1\Delta_{4}}&=&-\partial_{v}\bar{D}_{\Delta_{1}\Delta_{2}\Delta_{3}\Delta_{4}}
\nonumber \\
\bar{D}_{\Delta_{1}\Delta_{2}\Delta_{3}+1\Delta_{4}+1}&=&(\Delta_{3}
+\Delta_{4}-\Sigma-u\partial_{u})\bar{D}_{\Delta_{1}\Delta_{2}\Delta_{3}\Delta_{4}}
\nonumber \\
\bar{D}_{\Delta_{1}+1\Delta_{2}\Delta_{3}\Delta_{4}+1}&=&(\Delta_{1}
+\Delta_{4}-\Sigma-v\partial_{v})\bar{D}_{\Delta_{1}\Delta_{2}\Delta_{3}\Delta_{4}}
\nonumber \\
\bar{D}_{\Delta_{1}\Delta_{2}+1\Delta_{3}\Delta_{4}+1}&=&(\Delta_{2}
+u\partial_{u}+v\partial_{v})\bar{D}_{\Delta_{1}\Delta_{2}\Delta_{3}\Delta_{4}}
\nonumber \\
\bar{D}_{\Delta_{1}+1\Delta_{2}\Delta_{3}+1\Delta_{4}}&=&(\Sigma-\Delta_{4}
+u\partial_{u}+v\partial_{v})\bar{D}_{\Delta_{1}\Delta_{2}\Delta_{3}\Delta_{4}}
\label{Dids1}
\end{eqnarray}
There are additional identities which relate $\bar{D}$-functions with different values of $\Sigma$, and can be derived by repeated use of $(\ref{Dids1})$. These are
\begin{eqnarray}
(\Delta_{2}+\Delta_{4}-\Sigma)\bar{D}_{\Delta_{1}\Delta_{2}\Delta_{3}\Delta_{4}}
&=&\bar{D}_{\Delta_{1}\Delta_{2}+1\Delta_{3}\Delta_{4}+1}
-\bar{D}_{\Delta_{1}+1\Delta_{2}\Delta_{3}+1\Delta_{4}} \nonumber \\
(\Delta_{1}+\Delta_{4}-\Sigma)\bar{D}_{\Delta_{1}\Delta_{2}\Delta_{3}\Delta_{4}}
&=&\bar{D}_{\Delta_{1}+1\Delta_{2}\Delta_{3}\Delta_{4}+1}
-v\bar{D}_{\Delta_{1}\Delta_{2}+1\Delta_{3}+1\Delta_{4}} \nonumber \\
(\Delta_{3}+\Delta_{4}-\Sigma)\bar{D}_{\Delta_{1}\Delta_{2}\Delta_{3}\Delta_{4}}
&=&\bar{D}_{\Delta_{1}\Delta_{2}\Delta_{3}+1\Delta_{4}+1}
-u\bar{D}_{\Delta_{1}+1\Delta_{2}+1\Delta_{3}\Delta_{4}}
\end{eqnarray}
Furthermore, there are identities relating $\bar{D}$-functions with the same $\Sigma$. The most frequently used is
\begin{equation}
\Delta_{4}\bar{D}_{\Delta_{1}\Delta_{2}\Delta_{3}\Delta_{4}}=\bar{D}_{\Delta_{1}\Delta_{2}\Delta_{3}+1\Delta_{4}+1}
+\bar{D}_{\Delta_{1}\Delta_{2}+1\Delta_{3}\Delta_{4}+1}
+\bar{D}_{\Delta_{1}+1\Delta_{2}\Delta_{3}\Delta_{4}+1}
\end{equation}
Finally, we comment on the various symmetries that these functions exhibit. By means of conformal symmetry, one can see that
\begin{eqnarray}
\bar{D}_{\Delta_{1}\Delta_{2}\Delta_{3}\Delta_{4}}(u,v)
&=&v^{-\Delta_{2}}\bar{D}_{\Delta_{1}\Delta_{2}\Delta_{4}\Delta_{3}}(u/v,1/v)
\nonumber \\
\bar{D}_{\Delta_{1}\Delta_{2}\Delta_{3}\Delta_{4}}(u,v)
&=&v^{\Delta_{4}-\Sigma}\bar{D}_{\Delta_{2}\Delta_{1}\Delta_{3}\Delta_{4}}(u/v,1/v)
\nonumber \\
\bar{D}_{\Delta_{1}\Delta_{2}\Delta_{3}\Delta_{4}}(u,v)
&=&v^{\Delta_{1}+\Delta_{4}-\Sigma}\bar{D}_{\Delta_{2}\Delta_{1}\Delta_{4}\Delta_{3}}(u,v)
\nonumber \\
\bar{D}_{\Delta_{1}\Delta_{2}\Delta_{3}\Delta_{4}}(u,v)
&=&u^{\Delta_{3}+\Delta_{4}-\Sigma}\bar{D}_{\Delta_{4}\Delta_{3}\Delta_{2}\Delta_{1}}(u,v)
\nonumber \\
\bar{D}_{\Delta_{1}\Delta_{2}\Delta_{3}\Delta_{4}}(u,v)
&=&\bar{D}_{\Delta_{3}\Delta_{2}\Delta_{1}\Delta_{4}}(v,u)
\nonumber \\
\bar{D}_{\Delta_{1}\Delta_{2}\Delta_{3}\Delta_{4}}(u,v)
&=&\bar{D}_{\Sigma-\Delta_{3}\Sigma-\Delta_{4}\Sigma-\Delta_{1}\Sigma-\Delta_{2}}(u,v)
\end{eqnarray}
\section{Exchange Diagrams}
We review here the various methods for computing exchange diagrams that are relevant to the calculation of the four-point supergravity amplitude. The various details we have omitted here can be found in \cite{D'Hoker:1999ni,Arutyunov:2002fh,Berdichevsky:2007xd}. The basic idea is to use the underlying symmetries of AdS space to write down an ansatz for the $z$-integral, and then use the Green function equation to determine the explicit functional dependence. As usual, we work in Euclidean  $AdS_{d+1}$  space with Poincar\'{e} coordinates 
\begin{equation}
ds^2=\frac{1}{z_0^2}(dz_0^2+dz^idz^i)
\end{equation}
Covariant derivatives involve the Levi-Civita connection, so the explicit form of the Christoffel symbols is also required
\begin{equation}
\label{chffelAdS}
\Gamma_{\mu\nu}^{\rho}=\frac{1}{z_0}(\delta^{\rho}_{0}\delta_{\mu\nu}-\delta^{\rho}_{\nu}\delta_{\mu 0}-\delta^{\rho}_{\mu}\delta_{\nu0})
\end{equation}
\subsection{Scalar Exchanges}
The scalar exchange integrals have been computed in \cite{D'Hoker:1999ni}. For our case, we only need to consider exchanges of chiral
primaries of weights $2$ and $n$, for the $s$ and $t$ channels, respectively. The generic exchange integral has
the form
\begin{equation}
A(w,\vec{x}_{1},\vec{x}_{2})=\int [dz] G_{\Delta}(z,w)\tilde{K}_{\Delta_{1}}(z,\vec{x}_{1})\tilde{K}_{\Delta_{2}}(z,\vec{x}_{2})
\label{scalarex}
\end{equation}
where $\Delta$ is the conformal weight of the exchanged scalar and $\tilde{K}_{\Delta}(z,x)$ is the unit normalized bulk-to-boundary scalar propagator introduced in (\ref{Kpropapp}).
The exchange integral transforms under inversion $z_{\mu}= z_{\mu}'/(z')^2$ as
\begin{equation}
A(w,\vec{x}_{1},\vec{x}_{2})=|\vec{x}_{12}|^{-2\Delta_{2}}I(w'-\vec{x}_{12}')
\end{equation}
where $I(w)$ is a function given by
\begin{equation}
I(w)=\int [dz] G_{\Delta}(z,w)z_{0}^{\Delta_{1}}\left(\frac{z_{0}}{z^2}\right)^{\Delta_{2}}
\end{equation}
which is invariant under scale transformations and under the Poincar\'{e} subgroup of $SO(5,1)$. This implies
that one can make the ansatz
\begin{equation}
I(w)=(w_{0})^{\Delta_{12}}f(t)
\end{equation}
where $\Delta_{12}=\Delta_{1}-\Delta_{2}$ and $t=w_{0}^2/w^2$. To determine the function $f(t)$, one uses the equation of motion for the
Green function $G_{\Delta}(z,w)$, which leads to a second order differential equation which can be explicitly solved. For the cases
here considered, it suffices to quote the results. In the $s$-channel, $\Delta_{1}=\Delta_{2}=\Delta=2$ and $m_{2}^2=-4$. Hence (\ref{scalarex}) becomes
\begin{equation}
A(w,\vec{x}_{1},\vec{x}_{2})=\frac{1}{4}|\vec{x}_{12}|^{-2}\tilde{K}_{1}(w,\vec{x}_{1})\tilde{K}_{1}(w,\vec{x}_{2})
\end{equation}
In the $t$-channel, $\Delta_{1}=2$, $\Delta_{3}=\Delta=n$ and $m_{n}^{2}=n(n-4)$. The $z$-integral (\ref{scalarex})  gives
\begin{equation}
A(w,\vec{x}_{1},\vec{x}_{3})=\frac{1}{4(n-1)}|\vec{x}_{13}|^{-2}\tilde{K}_{1}(w,\vec{x}_{1})\tilde{K}_{n-1}(w,\vec{x}_{3})
\end{equation}
\subsection{Vector Exchanges}
The $z$-integrals for massless and massive vector exchanges have been computed before \cite{D'Hoker:1999ni}. We will just use the results and adapt them to our case. One is interested in diagrams of the form
\begin{equation}
A_{\mu}(w,\vec{x}_{1},\vec{x}_{2})=\int [dz] G_{\mu\nu'}(z,w)g^{\nu'\rho'}(z)\tilde{K}_{\Delta_{1}}(z,\vec{x}_{1})
\frac{\buildrel\leftrightarrow\over\partial}{\partial z_{\rho'}}\tilde{K}_{\Delta_{2}}(z,\vec{x}_{2})
\end{equation}
with $\tilde{K}_{\Delta}(z,\vec{x})$ as before. The propagator transforms as a bitensor under inversion, so when going to the inverted frame the expression above becomes
\begin{equation}
A_{\mu}(w,\vec{x}_{1},\vec{x}_{2})=|\vec{x}_{12}|^{-2\Delta_{2}}\frac{J_{\mu\nu}(w)}{w^2}I_{\nu}(w'-\vec{x}_{12}')
\label{vectorex}
\end{equation}
where $J_{\mu\nu}(w)=\delta_{\mu\nu}-2w_{\mu}w_{\nu}/w^2$ is the conformal jacobian and
\begin{equation}
I_{\mu}(w)=\int [dz] {G_{\mu}}^{\nu'}(z,w)z_{0}^{\Delta_{1}}\frac{\buildrel\leftrightarrow\over\partial}{\partial z_{\nu'}}
\left(\frac{z_{0}}{z^2}\right)^{\Delta_{2}}
\end{equation}
Using scale and Poincar\'{e} symmetries, one can write an ansatz for this integral
\begin{equation}
I_{\mu}(w)=w_{0}^{\Delta_{12}}\frac{w_{\mu}}{w^2}f(t)+w_{0}^{\Delta_{12}}\frac{\delta_{\mu 0}}{w_{0}}h(t)
\end{equation}
In order to determine the functions $f(t)$ and $h(t)$, one uses the corresponding Green function equation. This gives a second order differential equations that can be solved in any case which involves fields from type IIB supergravity compactified in $AdS_{5}\times S^{5}$.
We refer the reader to the formulas in \cite{D'Hoker:1999ni} that determine the solutions to this system. Note that in the case in which $\Delta_{1}=\Delta_{2}$, the solution is even simpler given that $h(t)=0$. We now write down the explicit results that interest us. For the $s$-channel, $\Delta_{1}=\Delta_{2}=2$ and $m^2=0$ so that $f(t)=\frac{1}{2}t$ and (\ref{vectorex}) becomes
\begin{eqnarray}
A_{\mu}(w,\vec{x}_{1},\vec{x}_{2})&=&\frac{|\vec{x}_{12}|^{-4}}{2}\frac{J_{\mu\nu}(w)}{w^2}
\left\{\frac{(w'-\vec{x}_{12}')_{\nu}}{(w'-\vec{x}_{12}')^2}\frac{w_{0}'}{(w'-\vec{x}_{12}')^2}\right\} \nonumber \\
&=&\frac{1}{2}\frac{1}{|\vec{x}_{12}|^2}\left\{\frac{(w-\vec{x}_{2})_{\mu}}{w_{0}}\tilde{K}_{2}(w,\vec{x}_{2})\tilde{K}_{1}(w,\vec{x}_{1})-
\frac{(w-\vec{x}_{1})_{\mu}}{w_{0}}\tilde{K}_{2}(w,\vec{x}_{1})\tilde{K}_{1}(w,\vec{x}_{2})\right\} \nonumber \\
\end{eqnarray}
For the $t$-channel, $\Delta_{1}=2$, $\Delta_{3}=n$ and $m^2=n(n-2)$, so that $f(t)=a_{1}t$ and $h(t)=b_{1}t$. One then gets
\begin{eqnarray}
A_{\mu}(w,\vec{x}_{1},\vec{x}_{3})&=&|\vec{x}_{31}|^{-4}\frac{J_{\mu\nu}(w)}{w^2}
\left\{a_{1}\frac{(w'-\vec{x}_{31}')_{\nu}}{(w'-\vec{x}_{31}')^2} +b_{1}\frac{\delta_{\nu 0}}{w_{0}'}\right\} \frac{w_{0}'}{(w'-\vec{x}_{31}')^2} {w_{0}'}^{n-2}\nonumber \\
&=&\frac{1}{|\vec{x}_{13}|^2}\left\{ \frac{a_{1}+2b_{1}}{2(n-1)}D_{\mu}\tilde{K}_{n-1}(w,\vec{x}_{3})\tilde{K}_{1}(w,\vec{x}_{1})
-\frac{a_{1}}{2}D_{\mu}\tilde{K}_{1}(w,\vec{x}_{1})\tilde{K}_{n-1}(w,\vec{x}_{3})\right\} \nonumber \\
\end{eqnarray}
and in this case, $a_{1}=-1/n$ and $b_{1}=0$.
\subsection{Symmetric Tensor Exchanges}
We now turn to the tensor exchanges. Again, all the ingredients to carry out this computation can be found in the 
literature \cite{D'Hoker:1999ni,Berdichevsky:2007xd}, so here
we just introduce the necessary ones. The idea is very similar to the one in the previous cases. One needs to compute the $z$-integral
\begin{equation}
A_{\mu\nu}(w,\vec{x}_{1},\vec{x}_{2})=\int [dz] G_{\mu\nu\mu'\nu'}(z,w)T^{\mu'\nu'}(z,\vec{x}_{1},\vec{x}_{2})
\end{equation}
with the tensor $T^{\mu\nu}(z,\vec{x}_{1},\vec{x}_{2})$ being of the form
\begin{eqnarray}
T^{\mu\nu}(w,\vec{x}_{1},\vec{x}_{2})&=&\nabla^{(\mu}\tilde{K}_{\Delta_{1}}(z,\vec{x}_{1})\nabla^{\nu)}\tilde{K}_{\Delta_{2}}(z,\vec{x}_{2})
-\frac{1}{2}g^{\mu\nu}\left(\nabla^{\rho}\tilde{K}_{\Delta_{1}}(z,\vec{x}_{1})\nabla_{\rho}\tilde{K}_{\Delta_{2}}(z,\vec{x}_{2}))\right.
\nonumber \\
&+&\left.\frac{1}{2}(m_{\Delta_{1}}^{2}+m_{\Delta_{2}}^{2}-k(k+4))\tilde{K}_{\Delta_{1}}(z,\vec{x}_{1})\tilde{K}_{\Delta_{2}}(z,\vec{x}_{2})\right)
\end{eqnarray}
where $m_{\Delta}^2=\Delta(\Delta-4)$ and $k$ is the weight of the exchanged tensor, which for our case can be either $0$ (massless graviton) or $n-2$ (massive graviton). To solve the
$z$-integral, one again inverts the expression above
\begin{equation}
A_{\mu\nu}(w,\vec{x}_{1},\vec{x}_{2})=|\vec{x}_{12}|^{-2\Delta_{3}}\frac{J_{\mu\lambda}(w)}{w^2}\frac{J_{\nu\rho}(w)}{w^2}I_{\lambda\rho}(w'-\vec{x}_{12}')
\end{equation}
and writes down an ansatz for this integral, guided by the existing symmetries. The most general ansatz has the form
\begin{equation}
I_{\mu\nu}(w)=w_{0}^{\Delta_{12}}g_{\mu\nu}h(t)+w_{0}^{\Delta_{12}}P_{\mu}P_{\nu}\phi(t)+w_{0}^{\Delta_{12}}\nabla_{\mu}\nabla_{\nu}X(t)
+2w_{0}^{\Delta_{12}}\nabla_{(\mu}(P_{\nu)}Y(t))
\end{equation}
where $P_{\mu}=\delta_{\mu 0}/w_{0}$ and $h(t)$, $\phi(t)$, $X(t)$, $Y(t)$ are undetermined functions. Here we should point out that in the
case in which $\Delta_{1}=\Delta_{2}$, the last two terms are pure diffeomorphisms and depend on the gauge choice of the propagator, so they
are left undetermined and do not have any physical effect, given that they drop out of the final $w$-integral.

In the $s$-channel amplitude, the process involves the exchange of a massless graviton, and the $z$-integral involves a vertex with two chiral primaries of weight 2. This integral has been worked in \cite{D'Hoker:1999ni}, so it suffices to present the final result. Here $\Delta_{1}=\Delta_{2}=2$, $m_{\Delta_{1}}^2=m_{\Delta_{2}}^2=-4$ and $k=0$. Hence
\begin{equation}
I_{\mu\nu}(w)=\frac{t}{3}\left\{ g_{\mu\nu}-3 P_{\mu}P_{\nu} \right\}
\end{equation}
For the $t$-channel, $\Delta_{1}=2$, $\Delta_{3}=n$, $m_{\Delta_{1}}^2=-4$, $m_{\Delta_{3}}^2=n(n-4)$ and $k=n-2$. Using manipulations such as the ones presented in \cite{Arutyunov:2002fh} and \cite{Berdichevsky:2007xd}, one can simplify the result to the expression
\begin{equation}
I_{\mu\nu}(w)=-{w_{0}}^{n-2}\frac{8nt}{(n+1)(n+2)}\frac{w_{\mu}w_{\nu}}{w^4}
\end{equation}
and one can rewrite both expressions in terms of the original coordinates. Note that
\begin{eqnarray}
{w_{0}}'&\rightarrow& \tilde{K}_{1}(w,\vec{x}) \nonumber \\
t=\tilde{K}_{1}(w,\vec{x}_{ij}') &\rightarrow& |\vec{x}_{ij}'|^2 \tilde{K}_{1}(w,\vec{x}_{i})\tilde{K}_{1}(w,\vec{x}_{j}) \nonumber \\
\frac{J_{\mu\lambda}(w)}{w^2}\frac{(w'-\vec{x}_{ij}')_{\lambda}}{(w'-\vec{x}_{ij}')^2} &\rightarrow&
Q_{\mu}=\frac{(w-\vec{x}_{i}')_{\mu}}{(w-\vec{x}_{i}')^2}-\frac{(w-\vec{x}_{j}')_{\mu}}{(w-\vec{x}_{j}')^2} \nonumber \\
\frac{J_{\mu\lambda}(w)}{w^2}P_{\mu}' &\rightarrow& R_{\mu}=P_{\mu}-2\frac{(w-\vec{x}')_{\mu}}{(w-\vec{x}')^2}
\end{eqnarray}
so the $z$-integrals in the original coordinates read
\begin{equation}
A_{\mu\nu}(w,\vec{x}_{1},\vec{x}_{2})=\frac{1}{3}\frac{1}{|\vec{x}_{12}|^2}\left\{g_{\mu\nu}
-3\left(P_{\mu}-2\frac{(w-\vec{x}_{1})_{\mu}}{(w-\vec{x}_{1})^2}\right)\left(P_{\nu}-2\frac{(w-\vec{x}_{2})_{\nu}}{(w-\vec{x}_{2})^2}\right)\right\}
\tilde{K}_{1}(w,\vec{x}_{1})\tilde{K}_{1}(w,\vec{x}_{2})
\end{equation}
for the $s$-channel amplitude and
\begin{equation}
A_{\mu\nu}(w,\vec{x}_{1},\vec{x}_{3})=-\frac{8n}{(n+1)(n+2)}\frac{1}{|\vec{x}_{13}|^2}Q_{\mu}Q_{\nu}\tilde{K}_{n-1}(w,\vec{x}_{3})\tilde{K}_{1}(w,\vec{x}_{1})
\end{equation}
for the $t$-channel amplitude.
\section{Reduction of Quartic Couplings}
\label{sec:QuarticSimp}
The calculation follows in the same lines as in \cite{Berdichevsky:2007xd} for the case in which $n=3$\footnote{We thank G. Arutyunov for providing the expressions of the couplings as the paper \cite{Arutyunov:1999fb} has some typos.}. One starts from the quartic lagrangian \cite{Arutyunov:1999fb}
\begin{eqnarray}
\mathcal{L}_{4}&=&\mathcal{L}_{k_{1}k_{2}k_{3}k_{4}}^{(0)I_{1}I_{2}I_{3}I_{4}}s_{k_{1}}^{I_{1}}s_{k_{2}}^{I_{2}}s_{k_{3}}^{I_{3}}s_{k_{4}}^{I_{4}}+
\mathcal{L}_{k_{1}k_{2}k_{3}k_{4}}^{(2)I_{1}I_{2}I_{3}I_{4}}s_{k_{1}}^{I_{1}}\nabla_{\mu}s_{k_{2}}^{I_{2}}s_{k_{3}}^{I_{3}}\nabla^{\mu}s_{k_{4}}^{I_{4}}
\nonumber \\
&+&\mathcal{L}_{k_{1}k_{2}k_{3}k_{4}}^{(4)I_{1}I_{2}I_{3}I_{4}}s_{k_{1}}^{I_{1}}\nabla_{\mu}s_{k_{2}}^{I_{2}}\nabla^{\nu}\nabla_{\nu}(s_{k_{3}}^{I_{3}}\nabla^{\mu}s_{k_{4}}^{I_{4}})
\end{eqnarray}
We use the formulas in appendix B to expand the products of $SO(6)$ tensors. We first consider the four-derivative couplings. There are six terms, one for each permutation of the $k_{i}$'s. One can find that there are two tensor structures that enter the expressions. These are given by
\begin{eqnarray}
A^{1234}&=& A_{1} \left( C^{1234} - C^{1243} \right)+A_{2}\left( \Upsilon^{1234} - \Upsilon^{1243} \right)
\nonumber \\
A_{1}&=&\frac{\left( -2 + n \right) \,\left( 6 + n \right) \,\left( -16 + 16\,n + n^2 \right) \,
    }{4096\,
    \left( -6 - 5\,n + 5\,n^2 + 5\,n^3 + n^4 \right) }
\nonumber \\
A_{2}&=&\frac{(n+2)}{65536\,n\,\left( -1 + n^2 \right) }\left[\frac{4\,\left( -1 + n \right) \,\left( 1 + n
\right) \,\left( 4 + n \right) \,
    \left( 528 + 368\,n + 140\,n^2 + 24\,n^3 + 3\,n^4 \right) \,{n!}^4}{{\left( -1 + n \right) !}^2\,{\left( 3 + n \right) !}^2}\right.
\nonumber \\
&+&\frac{n\,{\left( 4 + n \right) }^2\,\left( 6 + n \right) \,\left( -1 + n^2 \right) \,
    \left( 144 + 96\,n + 44\,n^2 + 8\,n^3 + 3\,n^4 \right) \,{\left( 2 + n \right) !}^2}{{\left( 2 + n \right) }^2\,
    {\left( 4 + n \right) !}^2}
\nonumber \\
&+&\frac{2\,{\sqrt{6}}\,\left( -2 + n \right) \,{\sqrt{\left( -1 + n \right) \,n}}\,\left( 1 + n \right) \,
    \left( 720 + 640\,n + 236\,n^2 + 40\,n^3 + 3\,n^4 \right) \,{n!}^2}{\left( \frac{3 - n}{2} \right) !\,
    \left( \frac{-1 + n}{2} \right) !\,\left( \frac{1 + n}{2} \right) !\,\left( 2 + n \right) !\,\left( \frac{7 + n}{2} \right) !
    }
\nonumber \\
&-&\left.\frac{9600\,\left( 1 + n \right) \,{n!}^2}{\left( 2 + n \right) \,\left( -2 + n \right) !\,\left( 3 + n
\right) !}\right]
\end{eqnarray}
and
\begin{equation}
\Sigma^{1234}=\frac{7\,{\left( -2 + n \right) }^2\,{\left( 6 + n \right) }^2\left[2\,\delta_{2}^{12}\,\delta_{n}^{34} +
  n\,\left( 4\,C^{1234} + 4\,C^{1243} +
     \left( -1 + n \right) \,\left( 4\,S^{1234} + \Upsilon^{1234} +
        \Upsilon^{1243} \right)  \right)\right]}
  {32768\,\left( -1 + n \right) \,n\,\left( 1 + n \right) \,\left( 2 + n \right) \,\left( 3 + n \right) \,\left( 4 + n \right) }
\end{equation}
Here $A^{1234}$ is antisymmetric under $3\leftrightarrow 4$ and $\Sigma^{1234}$ is symmetric. One can reduce the four-derivative term by using the following formula
\begin{eqnarray}
s_{k_{1}}^{1}\nabla_{\mu}s_{k_{2}}^{2}\nabla &\cdot& \nabla ( s_{k_{3}}^{3}\nabla^{\mu}s_{k_{4}}^{4})=
\nonumber \\
&&(m_{k_{3}}^{2}+m_{k_{4}}^{2}-4)s_{k_{1}}^{1}\nabla_{\mu}s_{k_{2}}^{2}s_{k_{3}}^{3}\nabla^{\mu}s_{k_{4}}^{4} +
2s_{k_{1}}^{1}\nabla_{\mu}s_{k_{2}}^{2}\nabla_{\nu}s_{k_{3}}^{3}\nabla^{\nu}\nabla^{\mu}s_{k_{4}}^{4}
\end{eqnarray}
Using this identity on each of the six terms, and using the symmetries of the tensors $A^{1234}$ and $\Sigma^{1234}$ one can show the four-derivative terms vanish, with the remaining contribution being
\begin{equation}
\mathcal{L}^{(4)}_{4}=\Sigma^{1234}(m_{2}^{2}+m_{n}^{2}-4)\left(-2s_{2}^{1}\nabla_{\mu}s_{2}^{2}s_{n}^{3}\nabla^{\mu}s_{n}^{4}
+s_{2}^{1}\nabla_{\mu}s_{n}^{3}s_{2}^{2}\nabla^{\mu}s_{n}^{4}+s_{n}^{3}\nabla_{\mu}s_{2}^{1}s_{n}^{4}\nabla^{\mu}s_{2}^{2}\right)
\label{fourvanish}
\end{equation}
One can still simplify this expression further by employing integration by parts. For a general tensor
$\Omega^{1234}$, one has
\begin{equation}
\Omega^{1234}s_{2}^{1}\nabla_{\mu}s_{n}^{3}s_{2}^{2}\nabla^{\mu}s_{n}^{4}=
-(\Omega^{1243}+\Omega^{1234})s_{2}^{1}\nabla_{\mu}s_{2}^{2}s_{n}^{3}\nabla^{\mu}s_{n}^{4}
-m_{n}^{2}s_{2}^{1}s_{2}^{2}s_{n}^{3}s_{n}^{4} \label{eqintparts}
\end{equation}
so using this in eq. (\ref{fourvanish}) and relabeling appropriately, the final form of the contribution from the four-derivative terms is
\begin{equation}
\mathcal{L}^{(4)}_{4}=\Sigma^{1234}(m_{2}^{2}+m_{n}^{2}-4)\left(-6s_{2}^{1}\nabla_{\mu}s_{2}^{2}s_{n}^{3}\nabla^{\mu}s_{n}^{4}
-(m_{2}^{2}+m_{n}^{2})s_{2}^{1}s_{2}^{2}s_{n}^{3}s_{n}^{4}\right)
\end{equation}
so we see that the four-derivative couplings vanish and that the lagrangian relevant to the computation is of $\sigma$-model type. This gives futher evidence that the complete fourth order Lagrangian may share this feature.

We now move to the two-derivative couplings contribution. One proceeds on similar grounds, so one finds
\begin{equation}
\mathcal{L}_{4}^{(2)}=B^{1234}_{1}s_{2}^{1}\nabla_{\mu}s_{2}^{2}s_{n}^{3}\nabla^{\mu}s_{n}^{4}
+B^{1234}_{2}(s_{n}^{3}\nabla_{\mu}s_{2}^{1}s_{n}^{4}\nabla^{\mu}s_{2}^{2}+s_{2}^{1}\nabla_{\mu}s_{n}^{3}s_{2}^{2}\nabla^{\mu}s_{n}^{4})
\label{fourder}
\end{equation}
where
\begin{eqnarray}
B_{1}^{1234}&=&\left(-4154598 - 9778848\,n + 2557080\,n^2 + 3842368\,n^3 + 2099672\,n^4\right.
\nonumber \\
&+& \left.747584\,n^5 + 72436\,n^6 - 15520\,n^7 - 2320\,n^8\right)\delta_{2}^{12}\delta_n^{34}
\nonumber \\
&+&\left(15766068\,n - 26840640\,n^2 - 24890064\,n^3 - 14042368\,n^4 - 6850832\,n^5 - 1800848\,n^6\right.
\nonumber \\
&-&\left. 306328\,n^7 - 51680\,n^8 - 4640\,n^9\right)C^{1234}
\nonumber \\
&+&\left(12227124\,n - 30674496\,n^2 - 22678224\,n^3 - 10355968\,n^4 - 5523728\,n^5 - 1653392\,n^6\right.
\nonumber \\
&-&\left. 306328\,n^7 - 51680\,n^8 - 4640\,n^9\right)C^{1243}
\nonumber \\
&+&\left(-73829940\,n + 63354228\,n^2 + 31719312\,n^3 - 7703120\,n^4 - 4855472\,n^5 - 5440560\,n^6\right.
\nonumber \\
&-&\left. 2625160\,n^7 - 546968\,n^8 - 67680\,n^9 - 4640\,n^{10}\right)S^{1234}
\nonumber \\
&+&\left(-21314445\,n + 7739229\,n^2 + 9468900\,n^3 + 2699116\,n^4 + 1986388\,n^5 - 45132\,n^6\right.
\nonumber \\
&-&\left. 397090\,n^7 - 118886\,n^8 - 16920\,n^9 - 1160\,n^{10}\right)(\Upsilon^{1234}+\Upsilon^{1243})
\nonumber \\
&/&\left(589824\,\left( -1 + n \right) \,n\,\left( 1 + n \right) \,\left( 2 + n \right) \,\left( 3 + n \right)
\,\left( 4 + n \right)\right)
\end{eqnarray}

\begin{eqnarray}
B_{2}^{1234}&=&\left(-2412774 - 10069152\,n + 1661976\,n^2 + 4035904\,n^3 + 2172248\,n^4\right.
\nonumber \\
&+&\left.741536\,n^5 + 70924\,n^6 - 15520\,n^7 - 2320\,n^8\right)\delta_{2}^{12}\delta_{n}^{34}
\nonumber \\
&+&\left(7167540\,n - 39344448\,n^2 - 10705872\,n^3 + 1528448\,n^4 + 364720\,n^5 - 32576\,n^6\right.
\nonumber \\
&-&\left. 127336\,n^7 - 48512\,n^8 - 4640\,n^9\right)C^{1234}
\nonumber \\
&+&\left(24254004\,n - 23165760\,n^2 - 38230992\,n^3 - 21466240\,n^4 - 11121872\,n^5 - 3298400\,n^6\right.
\nonumber \\
&-&\left. 491368\,n^7 - 54848\,n^8 - 4640\,n^9\right)C^{1243}
\nonumber \\
&+&\left(-73774644\,n + 71252340\,n^2 + 30717072\,n^3 - 13566800\,n^4 - 5940656\,n^5 - 5430768\,n^6\right.
\nonumber \\
&-&\left. 2634232\,n^7 - 549992\,n^8 - 67680\,n^9 - 4640\,n^{10}\right)S^{1234}
\nonumber \\
&+&\left(-31286157\,n - 6479139\,n^2 + 22595748\,n^3 + 9096940\,n^4 + 5602612\,n^5 + 1044780\,n^6\right.
\nonumber \\
&-&\left. 405958\,n^7 - 149162\,n^8 - 18504\,n^9 - 1160\,n^{10}\right)\Upsilon^{1234}
\nonumber \\
&+&\left(-13084557\,n + 23989725\,n^2 - 3053148\,n^3 - 4787348\,n^4 - 1508876\,n^5 - 1056420\,n^6\right.
\nonumber \\
&-&\left. 392758\,n^7 - 90122\,n^8 - 15336\,n^9 - 1160\,n^{10}\right)\Upsilon^{1243}
\nonumber \\
&/&\left(2359296\,\left( -1 + n \right) \,n\,\left( 1 + n \right) \,\left( 2 + n \right) \,\left( 3 + n \right)
\,\left( 4 + n \right)\right)
\end{eqnarray}
One can again use eq. (\ref{eqintparts}) to rewrite this as
\begin{equation}
\mathcal{L}_{4}^{(2)}=\tilde{B}^{1234}_{1}s_{2}^{1}\nabla_{\mu}s_{2}^{2}s_{n}^{3}\nabla^{\mu}s_{n}^{4} -
(m_{2}^{2}+m_{n}^{2})B_{2}^{1234}s_{2}^{1}s_{2}^{2}s_{n}^{3}s_{n}^{4} \label{twoder}
\end{equation}
where
\begin{eqnarray}
\tilde{B}_{1}^{1234}&=&\left(-48384 + 8064\,n + 24864\,n^2 - 5376\,n^3 - 2016\,n^4 + 168\,n^5 +
42\,n^6\right)\delta_{2}^{12}\delta_{n}^{34}
\nonumber \\
&+&\left(1536\,n + 122624\,n^2 - 11712\,n^3 - 113152\,n^4 - 40896\,n^5 - 3760\,n^6 + 84\,n^7\right)C^{1234}
\nonumber \\
&+&\left(-96768\,n + 16128\,n^2 + 49728\,n^3 - 10752\,n^4 - 4032\,n^5 + 336\,n^6 + 84\,n^7\right)C^{1243}
\nonumber \\
&+&\left(-1536\,n - 219392\,n^2 + 27840\,n^3 + 162880\,n^4 + 30144\,n^5 - 272\,n^6 + 252\,n^7 +
84\,n^8\right)S^{1234}
\nonumber \\
&+&\left(24192\,n - 28224\,n^2 - 8400\,n^3 + 15120\,n^4 - 1680\,n^5 - 1092\,n^6 + 63\,n^7 +
21\,n^8\right)\times
\nonumber \\
&&(\Upsilon^{1234}+\Upsilon^{1234})
/\left(16384\,\left( -1 + n \right) \,n\,\left( 1 + n \right) \,\left( 2 + n \right) \,\left( 3 + n \right)
\,\left( 4 + n \right)\right) 
\end{eqnarray}
Finally we write down the contribution from the non-derivative terms. Using the symmetries $1 \leftrightarrow 2$
and $3 \leftrightarrow 4$, one gets
\begin{equation}
\mathcal{L}_{4}^{(0)}=C^{1234}_{1}s_{2}^{1}s_{2}^{2}s_{n}^{3}s_{n}^{4} \label{zeroder}
\end{equation}
where
\begin{eqnarray}
C^{1234}_{1}&=&\left[\left(7147980 + 26899212\,n + 16985757\,n^2 - 13435136\,n^3 - 13143972\,n^4 -
7471824\,n^5\right.\right.
\nonumber \\
&-& \left.1900172\,n^6 + 112504\,n^7 + 71646\,n^8 - 3120\,n^9 - 1160\,n^{10}\right)\delta^{12}_{2}\delta^{34}_{n}
\nonumber \\
&+&\left(-53553360\,n + 48685104\,n^2 + 203418228\,n^3 + 112576768\,n^4 + 63864432\,n^5\right.
\nonumber \\
&+&\left.29953152\,n^6 + 4595536\,n^7 - 73904\,n^8 - 82056\,n^9 - 33120\,n^{10} - 4640\,n^{11}\right)C^{1234}
\nonumber \\
&+&\left(142904424\,n + 19974528\,n^2 - 227914530\,n^3 - 21668294\,n^4 + 43279688\,n^5 + 22968216\,n^6\right.
\nonumber \\
&+&\left.16668088\,n^7 + 4026888\,n^8 - 82780\,n^9 - 129348\,n^{10} - 24560\,n^{11} - 2320\,n^{12}\right)S^{1234}
\nonumber \\
&+&\left(86418996\,n + 54107328\,n^2 - 88656273\,n^3 - 36437731\,n^4 - 10739036\,n^5 - 9066804\,n^6\right.
\nonumber \\
&+&\left.\left. 2398940\,n^7 + 1888452\,n^8 + 146386\,n^9 - 46818\,n^{10} - 12280\,n^{11} -
1160\,n^{12}\right)\Upsilon^{1234}\right]
\nonumber \\
&/&\left(1179648\,\left( -1 + n \right) \,n\,\left( 1 + n \right) \,\left( 2 + n \right) \,\left( 3 + n \right)
\,\left( 4 + n \right)\right)
\end{eqnarray}
so combining equations (\ref{fourder}), (\ref{twoder}) and (\ref{zeroder}), one gets the simple expression for the contribution from the quartic lagrangian in (\ref{quartic}).

\section*{Acknowledgements}
I am very grateful to H. Osborn for suggesting the problem and for bringing to my attention the use of harmonic polynomials for the evaluation of the effective couplings. I am also indebted to C. Rayson for allowing me to see his results prior to publication. I would also like to thank M.B. Green for enlightening discussions and his valuable comments on the contents of this paper. Finally, I would like to thank G. Arutyunov and L. Berdichevsky for useful comments, and M. Paulos for his help with Cadabra.

This work has been supported by CONACyT M\'{e}xico and the ORSAS UK Award Scheme.

\bibliographystyle{ieeetr}
\bibliography{morediffw}

\end{document}